\documentclass[10pt]{article}

\usepackage{authblk}

\usepackage{graphicx} 
\usepackage[scriptsize,nooneline,hang]{caption}
\usepackage[hang,nooneline,scriptsize]{subfigure}
\usepackage{pstricks,pstricks-add}
\usepackage{pst-plot}
\usepackage{color}

\usepackage{amsmath,amsfonts,bbm,dsfont,mathrsfs} 

\newcommand{\half}{\frac{1}{2}}
\newcommand{\dd}{{\mathrm{d}}}
\newcommand{\tr}{\tilde{r}}
\newcommand{\tlt}{\tilde{t}}
\newcommand{\tv}{\tilde{v}}
\newcommand{\tL}{\tilde{L}}
\newcommand{\tK}{\tilde{K}}
\newcommand{\ta}{\tilde{a}}
\newcommand{\tb}{\tilde{\beta}}
\newcommand{\tD}{\tilde{\Delta}}
\newcommand{\tS}{\tilde{\Sigma}}

\begin{document}

\title{Analytic solutions of the geodesic equation for Einstein-Maxwell-dilaton-axion black holes}

\author{Kai Flathmann}
\author{Saskia Grunau}
\affil{Institut f\"ur Physik, Universit\"at Oldenburg, D--26111 Oldenburg, Germany}

\maketitle

\begin{abstract}
In this article we study the geodesic motion of test particles and light in the Einstein-Maxwell-dilaton-axion black hole spacetime. We derive the equations of motion and present their solutions in terms of the Weierstra{\ss} $\wp$-, $\sigma$- and $\zeta$-functions. With the help of parametric diagrams and effective potentials we analyze the geodesic motion and give a list of all possible orbit types.
\end{abstract}

\section{Introduction}

The discoveries in the last decades revealed that  approximately 95\% of the Universe is made of dark matter and dark energy. Until today the true nature of dark matter and dark energy remains an open question. The string theory, one of the best candidates to solve the problem of quantum gravity, may provide an answer. Scalar fields, which arise in some superstring theories, are promising candidates to explain dark energy and also dark matter. The low energy limit of the bosonic sector of the heterotic string theory is called Einstein-Maxwell-dilaton-axion (EMDA) gravity. This model is a nontrivial generalization of Einstein-Maxwell gravity and contains two scalar fields: the dilaton and the axion, which can be related to dark energy and dark matter (see e.g. \cite{Matarrese:2011}). The four-dimensional EMDA model is described by the action
\begin{equation}
 S = \int\! \dd^4 x \sqrt{-g} \left( R - 2g^{\mu\nu} \partial_\mu\phi \partial_\nu\phi - \half \mathrm{e}^{4\phi} g^{\mu\nu} \partial_\mu\kappa \partial_\nu\kappa - \mathrm{e}^{-2\phi} F_{\mu\nu} F^{\mu\nu} - \kappa F_{\mu\nu} \check{F}^{\mu\nu}\right) \, ,
\end{equation}
where  $\phi$ is the scalar dilaton field and $\kappa$ is the axion field, which is dual to the antisymmetric tensor field ${H=- \mathrm{e}^{4\phi} \ast \dd\kappa / 4}$. ${ \check{F}_{\mu\nu} = -\half \sqrt{-g} \epsilon_{\mu\nu\alpha\beta} F^{\alpha\beta} }$ is the dual of the electromagnetic tensor. A black hole solution of the EMDA field equations resulting from this action was found by Garcia, Galtsov and Kechkin~\cite{Garcia:1995qz}.

A powerful tool to study black holes is geodesics, since the structure of the orbits of particles and light provides information about the black hole. The analytical solutions of the geodesic equations are useful to calculate the shadow of a black hole or observables like the periastron shift. Those could later be compared to observations to test models and theories.

The first to find the analytical solutions of the geodesic equations in a black hole spacetime was Hagihara \cite{Hagihara:1931}. He solved the equations of motion in the Schwarzschild spacetime in terms of the elliptic Weierstra{\ss} $\wp$-function. In the Taub-NUT \cite{Kagramanova:2010bk}, Reissner-Nordstr\"om \cite{Grunau:2010gd}, Myers-Perry (with equal rotation parameters) \cite{Kagramanova:2012hw} and Kerr-Newman \cite{Hackmann:2013pva} spacetimes, the equations of motion are also of elliptic type and were solved in the same way. 

However, if black holes in higher dimensions are considered or if the cosmological constant is included, the equations of motion are usually hyperelliptic. In a nutshell the integration of the geodesic equations can be traced back to the Jacobi inversion problem and the solutions are given in terms of the $\sigma$-function. This was done for the four-dimensional Schwarzschild-de Sitter spacetime \cite{Hackmann:2008zza,Hackmann:2008zz} as well as for higher-dimensional Schwarzschild, Schwarzschild-(anti)de Sitter, Reissner-Nordstr\"om and Reissner-Nordstr\"om -(anti) de Sitter spactime \cite{Hackmann:2008tu}. Likewise the geodesic equations were solved in the Kerr-(anti) de Sitter spacetime \cite{Hackmann:2010zz}, in the higher-dimensional Myers-Perry spacetime \cite{Enolski:2010if} and in the  Ho\v{r}ava-Lifshitz spactime~\cite{Enolski:2011id}.
In five-dimensional black ring spacetimes, the equations of motion could be solved analytically in special cases \cite{Grunau:2012ai,Grunau:2012ri}. Moreover, the motion of test particles was studied in various black string spacetimes \cite{Aliev:1988wv, Galtsov:1989ct, Chakraborty:1991mb, Ozdemir:2003km, Ozdemir:2004ne, Grunau:2013oca} including field theoretical cosmic string spacetimes \cite{Hartmann:2010rr, Hartmann:2012pj,Hartmann:2010vp} and black holes pierced by a black string \cite{Hackmann:2009rp, Hackmann:2010ir}.\\
\enlargethispage{\baselineskip}

In this article we study the geodesic motion in the EMDA black hole spacetime. We analyze the possible orbit types using effective potential techniques and parametric diagrams. Furthermore we present the analytical solutions of the equations of motion for test particles and light. The equations of motion are of elliptic type and the solutions are given in terms of the  Weierstra{\ss} $\wp$-, $\sigma$- and $\zeta$-functions.

\section{The Einstein-Maxwell-Dilaton-Axion black hole}
\label{sec:EMDA}

The metric of the stationary axisymmetric EMDA black hole is given by \cite{Garcia:1995qz}
\begin{equation}
	\begin{split}
		\dd s^2 = &-\frac{\Sigma-a^2\sin^2\!\vartheta}{\Delta} \dd t^2 + \frac{2a\sin^2\!\vartheta }{\Delta} \left[\left(r^2 -2\beta r +a^2 \right)-\Sigma\right] \dd t\dd\varphi + \frac{\Delta}{\Sigma}\dd r^2 + \Delta\dd\vartheta^2\\ 
		&+ \frac{\sin^2\!\vartheta}{\Delta} \left[ \left(r^2 -2\beta r +a^2 \right)^2 -\Sigma a^2\sin^2\!\vartheta \right] \dd \varphi^2 \, ,
	\end{split}
\end{equation}
with
\begin{equation}
	\begin{split}
		\Sigma & = r^2 - 2mr + a^2 \, , \\
		\Delta & = r^2 - 2\beta r + a^2\cos^2\!\vartheta \, .
	\end{split}
\end{equation}
In comparison with the original solution of \cite{Garcia:1995qz}, here we chose $b=0$. If the parameter $b$ is different from zero, the solution carries a NUT charge. 

The parameter $\beta$ is the dilaton charge, $m$ is related to the ADM mass $M=m-\beta$, and $a$ corresponds to the angular momentum $J=aM$. The electric charge of the black hole is $Q=\sqrt{-2\omega\beta M}$, where $\omega=\exp(2\phi_0)$ and $\phi_0$ is the asymptotic value of the dilaton field (see \cite{Garcia:1995qz}). Demanding that $Q$ is real and $M>0$, we can conclude that $\beta\leq0$. If $\beta=0$, the solution reduces to the Kerr spacetime.

The horizons are given by $\Sigma=0$ and can be written in terms of the ADM mass
\begin{equation}
	r_\pm = M+\beta \pm \sqrt{\left( M+\beta \right)^2 -a^2} \, .
\end{equation}
The condition $a^2\leq \left( M+\beta \right)^2$ has to be fulfilled in order to get real values of $r_\pm$. In contrast to the Kerr spacetime, here it is possible that $r_+$ and $r_-$ are negative. In this article we will only analyze the case of positive horizons in detail, but nonetheless the analytical solutions of the geodesic equations given in later sections are valid in all cases.

The ergoregion is determined by $\Sigma-a^2\sin^2\!\vartheta =0$, for example in the equatorial plane ($\vartheta=\frac{\pi}{2}$) the boundary of the ergoregion is at $r=2\left( M+\beta \right)$. Inside this region the test particles have to rotate in the same direction as the black hole.

\begin{figure}[h]
	\centering
	\subfigure[$a=0.1$ and $\beta=-0.08$]{
		\includegraphics[width=0.3\textwidth]{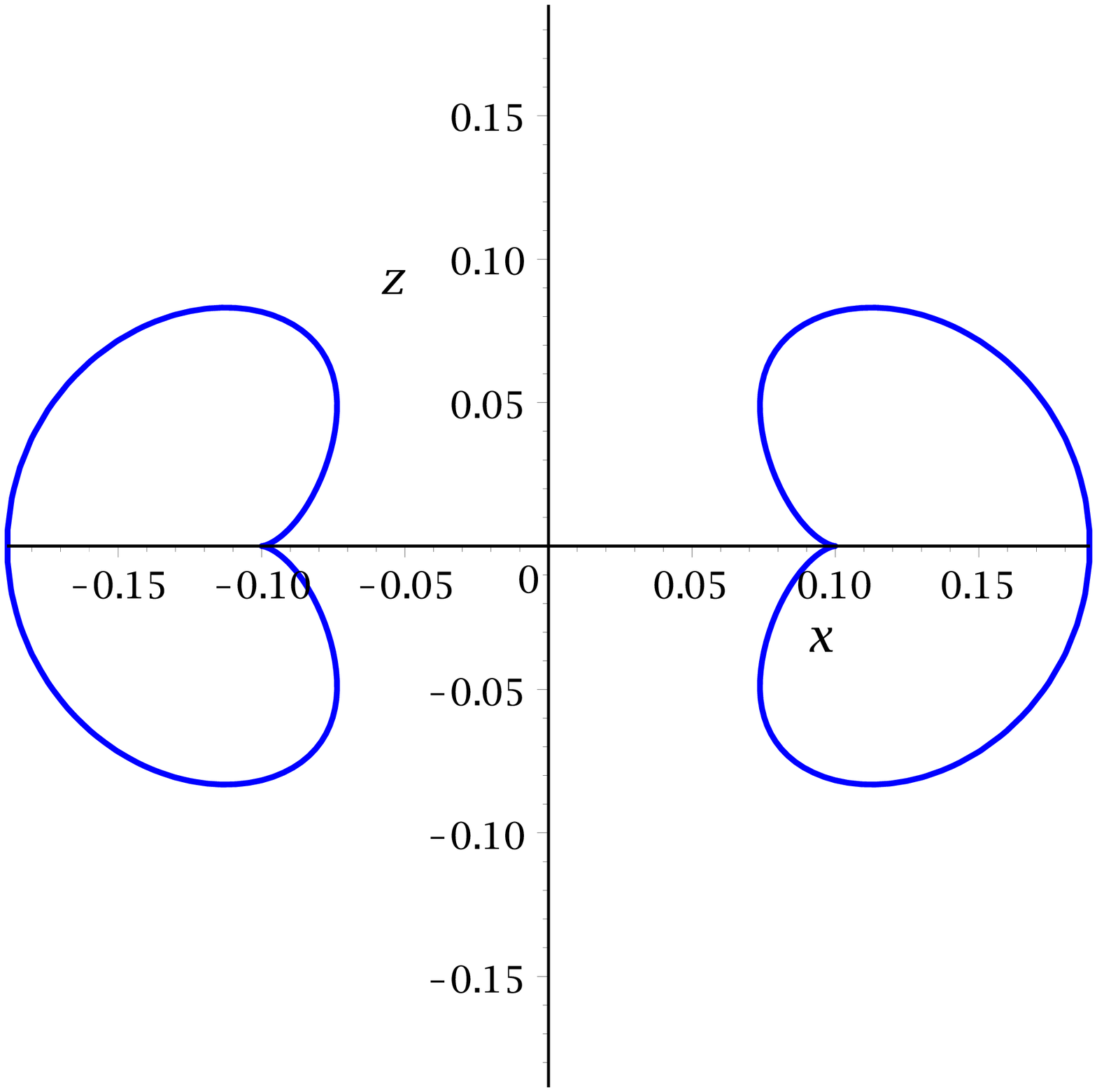}
	}
	\subfigure[$a=0.08$ and $\beta=-0.08$]{
		 \includegraphics[width=0.3\textwidth]{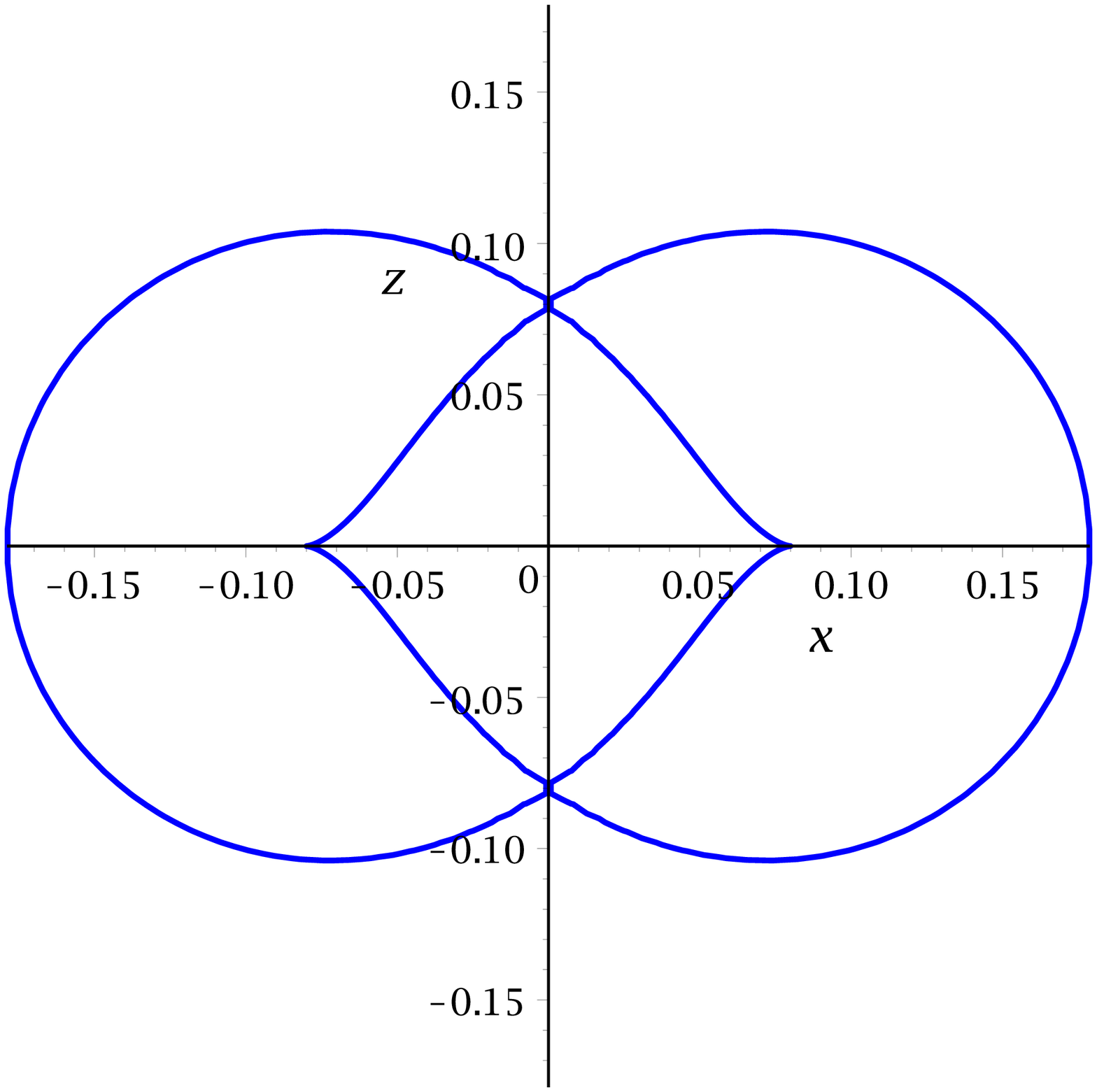}
	}
	\subfigure[$a=0.078$ and $\beta=-0.08$]{
		 \includegraphics[width=0.3\textwidth]{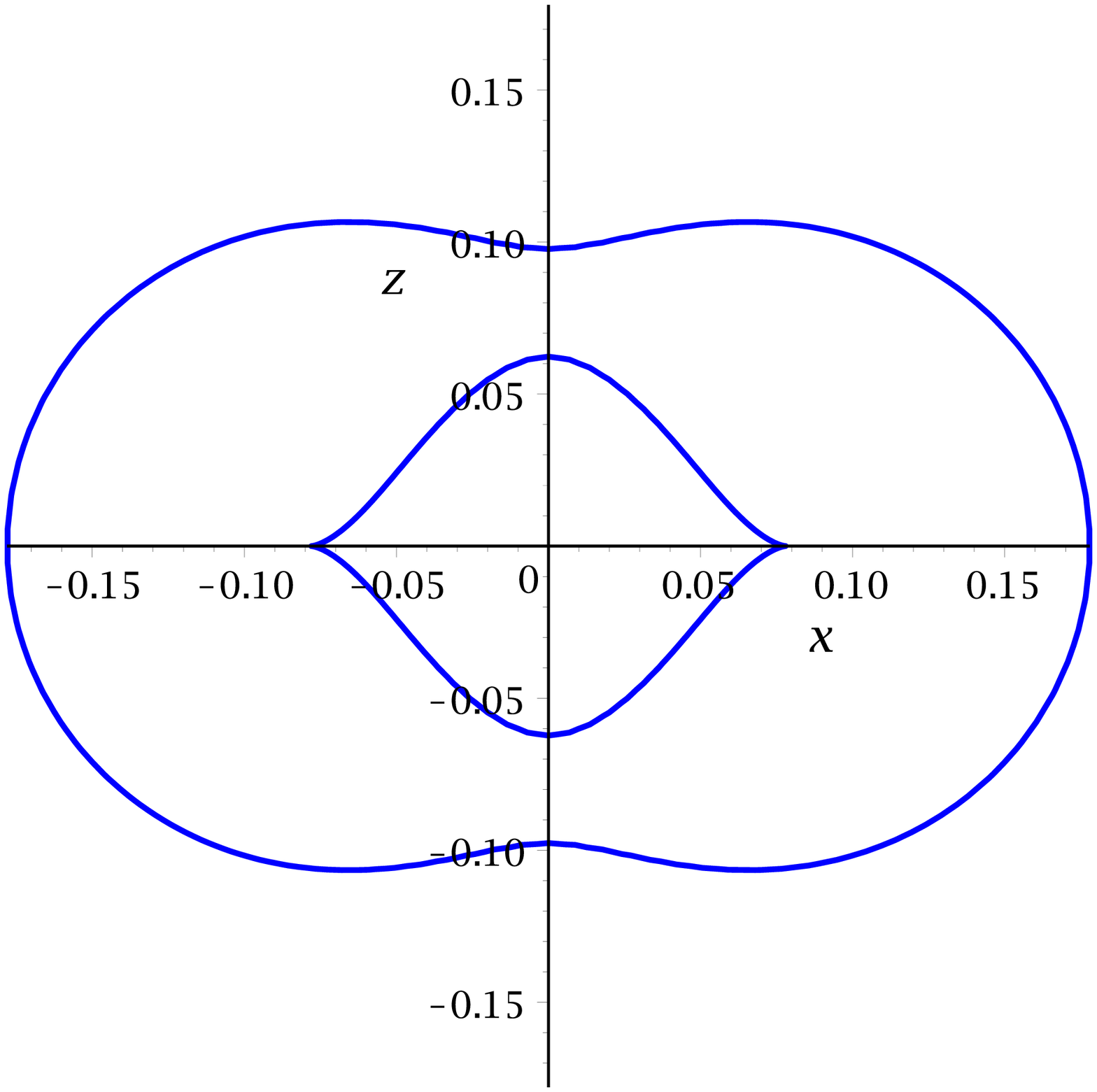}
	}
	\caption{Plots of the singularity given by $\Delta=0$, note that $r\leq 0$ in the figures.}
	\label{pic:singularity-shape}
\end{figure}

The singularity is obtained for $\Delta= r^2 - 2\beta r + a^2\cos^2\!\vartheta=0$, since the Kretschmann scalar diverges for $\Delta=0$. Due to the influence of the dilaton parameter $\beta$ the shape of the singularity is different from the ring singularity of a Kerr black hole. At $r=0$ and $\vartheta=\frac{\pi}{2}$ we still have the ring singularity, but for negative $r$ the singularity extends up to $r=2\beta$ (there is no singularity at positive $r$). Its shape varies from a deformed torus to a complex structure of two closed surfaces depending on the parameters $a$ and $\beta$, see figure~\ref{pic:singularity-shape} for a two-dimensional projection. However, only geodesics with $r=0$ and $\vartheta=\frac{\pi}{2}$ end in the singularity. Geodesics coming from positive $r$ can cross  $r=0$ if $\vartheta\neq\frac{\pi}{2}$, and continue to negative $r$. For negative $r$ values the geodesics enter a universe where gravity is repulsive (which can be seen in plots of the effective potential and the orbits in later sections); therefore orbits with  $r<0$ cannot hit the  singularity.\\

The metric is given in Boyer-Lindquist-like coordinates ($r$, $\vartheta$, $\varphi$), which can be transformed to Cartesian coordinates by
\begin{equation}
	\begin{split}
		x&=\sqrt{(r^2+a^2)}\sin\vartheta\cos\varphi \\
		y&=\sqrt{(r^2+a^2)}\sin\vartheta\sin\varphi \\
		z&=r\cos\vartheta \, .
	\end{split}
\end{equation}

\subsection{The geodesic equations}

The Hamilton-Jacobi equation
\begin{equation}
	\frac{\partial S}{\partial \tau} + \half g^{\mu\nu} \frac{\partial S}{\partial x^\mu}\frac{\partial S}{\partial x^\nu}=0
\label{eqn:ham-jac}
\end{equation}
can be solved with an ansatz for the action $S$
\begin{equation}
	S=\half\delta \tau -Et+L\varphi + S_r(r) + S_\vartheta(\vartheta) \, ,
\end{equation}
where $\tau$ is an affine parameter along the geodesic,  $E$ is the energy and $L$ is the angular momentum of the  test particle. The parameter $\delta$ represents the mass of the test particle and is equal to $1$ for a particle with nonzero mass and equal to $0$ for light.

The Hamilton-Jacobi equation \eqref{eqn:ham-jac} separates with the help of the Carter \cite{Carter:1968rr} constant $K$ and yields four differential equations
\begin{eqnarray}
	\left(\frac{\dd \tr}{\dd \gamma}\right) ^2 &=& R \, , \label{eqn:r-equation}\\
	\left(\frac{\dd \vartheta}{\dd \gamma}\right) ^2 &=& \Theta  \, , \label{eqn:theta-equation}\\
	\frac{\dd \varphi}{\dd \gamma} &=& \frac{\ta}{\tS} \left[\left(\tr^2-2\tb\tr+\ta^2\right)E-\ta\tL\right] + \frac{1}{\sin^2\!\vartheta} \left(\tL - \ta E \sin^2\!\vartheta \right)  \, , \label{eqn:phi-equation}\\
	\frac{\dd \tlt}{\dd \gamma} &=& \frac{\tr^2-2\tb\tr+\ta^2}{\tS} \left[\left(\tr^2-2\tb\tr+\ta^2\right)E-\ta\tL\right] + \ta \left(\tL - \ta E \sin^2\!\vartheta \right)  \, .  \label{eqn:t-equation}
\end{eqnarray}
The polynomial $R$ and the function $\Theta$ are given by
\begin{equation}
	\begin{split}
		R &=\left[\left(\tr^2-2\tb r+\ta^2\right)E-\ta\tL\right]^2 - \tS\left(\tK+\delta \left(\tr^2-2\tb\tr\right)\right) \, , \\
		\Theta &= \tK - \delta\ta^2\cos^2\!\vartheta - \frac{1}{\sin^2\vartheta} \left(\tL -\ta E \sin^2\!\vartheta\right) ^2  \, ,
	\end{split}
\end{equation}
with
\begin{equation}
	\begin{split}
		\tS & = \tr^2 - \left(1+ 2\tb\right) \tr + \ta^2 \, , \\
		\tD & = \tr^2 -2\tb \tr + \ta^2\cos^2\!\vartheta \, .
	\end{split}
\end{equation}
For convenience we introduced dimensionless quantities by scaling with the ADM mass
\begin{equation}
	\tr=\frac{r}{2M} \ , 
	\,\, \tlt=\frac{t}{2M} \ , 
	\,\, \tilde{\tau}=\frac{\tau}{2M} \ , 
	\,\, \tL=\frac{L}{2M} \ ,
	\,\, \ta=\frac{a}{2M} \ , 
	\,\, \tb=\frac{\beta}{2M} \ , 
	\,\, \tK=\frac{K}{2M}  \ .
\end{equation}
We also employed the Mino time \cite{Mino:2003yg} $\gamma$ as $\tD\dd\gamma = \dd\tilde{\tau}$.

\section{Classification of the geodesics}

The function $\Theta$ in equation \eqref{eqn:theta-equation} and the polynomial $R$ in equation \eqref{eqn:r-equation} define the characteristics of the geodesics. In this section we will therefore study the function $\Theta$ and the polynomial $R$ to determine the possible orbits of light and test particles in the EMDA spacetime. In \cite{ONeill:1995} a similar analysis was done for the Kerr spacetime; the Kerr–(anti–)de Sitter spacetime and the rotating black string spacetime were discussed analogously in \cite{Hackmann:2010zz} and \cite{Grunau:2013oca}, respectively.\\

Geodesic motion is possible if $\Theta\geq0$ and $R\geq0$, then real values of the coordinates $\vartheta$ and $\tr$ are obtained. From $\Theta\geq0$ it follows immediately that $\tK\geq0$. If $\tK=0$, the motion is confined to the equatorial plane ($\vartheta = \frac{\pi}{2}$) and $\tL=  \ta E$; otherwise an orbit is not possible \cite{ONeill:1995}.

As discussed before, a geodesic hits the ring singularity if $\tr=0$ and simultaneously $\vartheta=\frac{\pi}{2}$. In this case we have
\begin{equation}
	\begin{split}
		\Theta\left(\frac{\pi}{2}\right)&=\tK-(\ta E-\tL)^2\geq0 \quad \text{and} \\
		R(0)&=-\ta^2[\tK-(\ta E-\tL)^2] \geq0 \, ,
	\end{split}
	\label{eqn:R-Theta-0}
\end{equation}
which yields a condition for orbits ending in the singularity: $\tK=(\ta E-\tL)^2$. This orbit also resides in the equatorial plane. Furthermore equation \eqref{eqn:R-Theta-0} leads to the following conclusions:
\begin{enumerate}
	\item $\tK>(\ta E-\tL)^2$: The geodesics cross $\vartheta=\frac{\pi}{2}$ but do not cross $\tr=0$.
	\item $\tK<(\ta E-\tL)^2$: The geodesics do not cross $\vartheta=\frac{\pi}{2}$ but it is possible to cross $\tr=0$.
\end{enumerate}

\subsection{The $\vartheta$-motion}

The $\vartheta$-equation is not influenced by the dilaton charge and therefore is the same as in the Kerr black hole spacetime. Here we will briefly present the main results of \cite{ONeill:1995} and \cite{Hackmann:2010zz}.

First we substitute $\nu=\cos^2\vartheta$ (with $\nu\in[0,1]$) into the function $\Theta(\vartheta)$:
\begin{equation}
	\Theta(\nu)=\tK-\delta\ta^2\nu-\ta^2 E^2(1-\nu)+2\ta E\tL -\frac{\tL^2}{(1-\nu)} \, .
\end{equation}
Since the zeros of $\Theta$ are the turning points of the latitudinal motion, we are interested in the number of zeros in the interval $[0,1]$. This number changes if a zero crosses 0 or 1, or if a $\Theta$ has a double zero. From  $\Theta(\nu=0)=0$ and  $\Theta(\nu=1)=0$, which is only possible for $L=0$ due to the pole at $\nu=1$, we get two boundary cases of the $\vartheta$-motion:
\begin{equation}
	\begin{split}
		\tL&=\ta E\pm\sqrt{\tK} \qquad \text{and} \\
		\tK&=\delta\ta^2 \, \wedge \, \tL=0 \, .
	 \end{split}
	 \label{eqn:theta-border1}
\end{equation}

To remove the pole of $\Theta$ at $\nu=1$ we consider
\begin{equation}
	\Theta'(\nu)= \ta^2(\delta-E^2)\nu^2 + [2\ta E(\ta E-\tL)-\delta\ta^2-\tK]\nu + \tK-(\ta E-\tL)^2 \, ,
\end{equation}
where $\Theta(\nu)=\frac{1}{1-\nu}\Theta'(\nu)$. Double zeros fulfill the conditions
\begin{equation}
	\Theta'(\nu)=0 \quad \text{and} \quad \frac{\dd \Theta' (\nu)}{\dd\nu}=0 \, ,
\end{equation}
which gives 
\begin{equation}
	\tL=\frac{E\pm (\ta^2\delta-\tK)\sqrt{E^2-\delta)}}{2\ta \delta} \, .
	\label{eqn:theta-border2}
\end{equation}
Therefore double zeros only occur for $E^2\geq\delta$.\\

Parametric diagrams can be drawn with the help of the equations \eqref{eqn:theta-border1} and \eqref{eqn:theta-border2} (see figure \ref{pic:parametric-diagrams}). These diagrams reveal four regions with different numbers of zeros. In regions (a) and (d) orbits are not possible since $\Theta(\nu)<0$ for all $\nu\in[0,1]$. There is one real zero in the interval $[0,1]$ in regions (b), here $\tK>(\ta E-\tL)^2$ and the geodesics cross $\vartheta=\frac{\pi}{2}$. Two real zeros in the interval $[0,1]$ can be found in region (d), where $\tK<(\ta E-\tL)^2$ and $\vartheta=\frac{\pi}{2}$ is not crossed. Crossing $\tr=0$ is possible in region (d) but not in region (b).\\

Consider the special case $\tL=0$, in which $\Theta$ has a single zero at $\nu_0=\frac{\ta^2E^2-\tK}{\ta^2(E^2-\delta)}$. For $\tK=\delta\ta^2$ this zero is located at $\nu=1$ and for $\tK=\ta^2E^2$ it is at $\nu=0$.  For $\tK\neq\delta\ta^2$, there are two cases
\begin{enumerate}
	\item $\tK>\delta\ta^2$: $\nu_0\in [0,1]$ if $\tK<\ta^2E^2$ and $\Theta(\nu)>0$ for all $\nu\in[0,1]$ if $\tK>\ta^2E^2$.
	\item $\tK<\delta\ta^2$: $\nu_0\in [0,1]$ if $\tK>\ta^2E^2$ but $\Theta(\nu)<0$ for all $\nu\in[0,1]$ if $\tK<\ta^2E^2$.
\end{enumerate}

\subsection{The $\tr$-motion}

The polynomial $R$ determines the possible orbit types, since its zeros are the turning points of the geodesics. One can distinguish between \emph{bound orbits} moving between two turning points and \emph{escape orbits}, where the test particles approach the black hole, turn around at a certain point and then escape to infinity. So-called \emph{terminating orbits} end in the singularity at $\tr=0$ and $\vartheta=\frac{\pi}{2}$, here  $\tK=(\ta E-\tL)^2$ is required. Sometimes the test particles cross both horizons twice or even multiple times. Since these orbits enter another universe, they are called \emph{two-world} or \emph{many-world orbits}. Similar to the Kerr spacetime, the $\tr$-coordinate is allowed to take negative values. An orbit that crosses $\tr=0$ is called \emph{transit} or \emph{crossover orbit} \cite{Hackmann:2010zz}. Let us first introduce a list of all possible orbits (for $\tr_\pm > 0$) before analyzing the $\tr$-motion in detail.
\begin{enumerate}
	\item \textit{Transit orbit} (TrO) with range $\tr \in (-\infty, \infty)$.
	\item \textit{Escape orbit} (EO) with range $\tr \in [r_1, \infty)$ with $r_1>\tr_+$, or with range $\tr \in (-\infty, r_1]$ with  $r_1<0$.
	\item \textit{Two-world escape orbit} (TEO) with range $[r_1, \infty)$ where $0<r_1 < r_-$.
	\item \textit{Crossover two-world escape orbit} (CTEO) with range $[r_1, \infty)$ where $r_1 < 0$.
	\item \textit{Bound orbit} (BO) with range $\tr \in [r_1, r_2]$ with
	\begin{enumerate}
		\item $r_1, r_2  > r_+$ or 
		\item $ 0 < r_1, r_2 < r_-$ or
		\item $r_1, r_2 < 0$  if $\tb \neq 0$.
	\end{enumerate}
	\item \textit{Many-world bound orbit} (MBO) with range $\tr \in [r_1, r_2]$ where $0<r_1 \leq r_-$ and $r_2 \geq r_+$.
	\item \textit{Terminating orbit} (TO) with ranges either $\tr \in [0, \infty)$ or $\tr \in [0, r_1]$ with
	\begin{enumerate}
		\item $r_1\geq\tr_+$ or 
		\item $0<r_1<\tr_-$ or
		\item $r_1<0$  if $\tb \neq 0$.
	\end{enumerate}
\end{enumerate}
Note that the bound orbit with $r_1, r_2 < 0$ and the terminating orbit with $r_1<0$ can only exist in the EMDA spacetime with $\tb \neq 0$, but not in the Kerr spacetime.

\begin{figure}[h]
	\centering
	\subfigure[$\delta=1$, $\ta=0.4$, $\tK=2$, $\tb=-0.08$]{
		\includegraphics[width=0.31\textwidth]{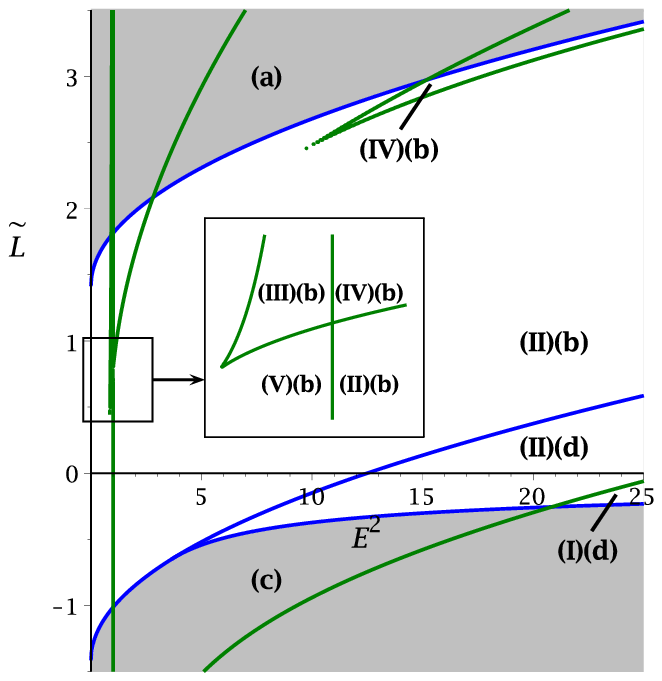}
	}
	\subfigure[$\delta=1$, $\ta=0.2$, $\tK=0.008$,  $\tb=-0.08$]{
		\includegraphics[width=0.31\textwidth]{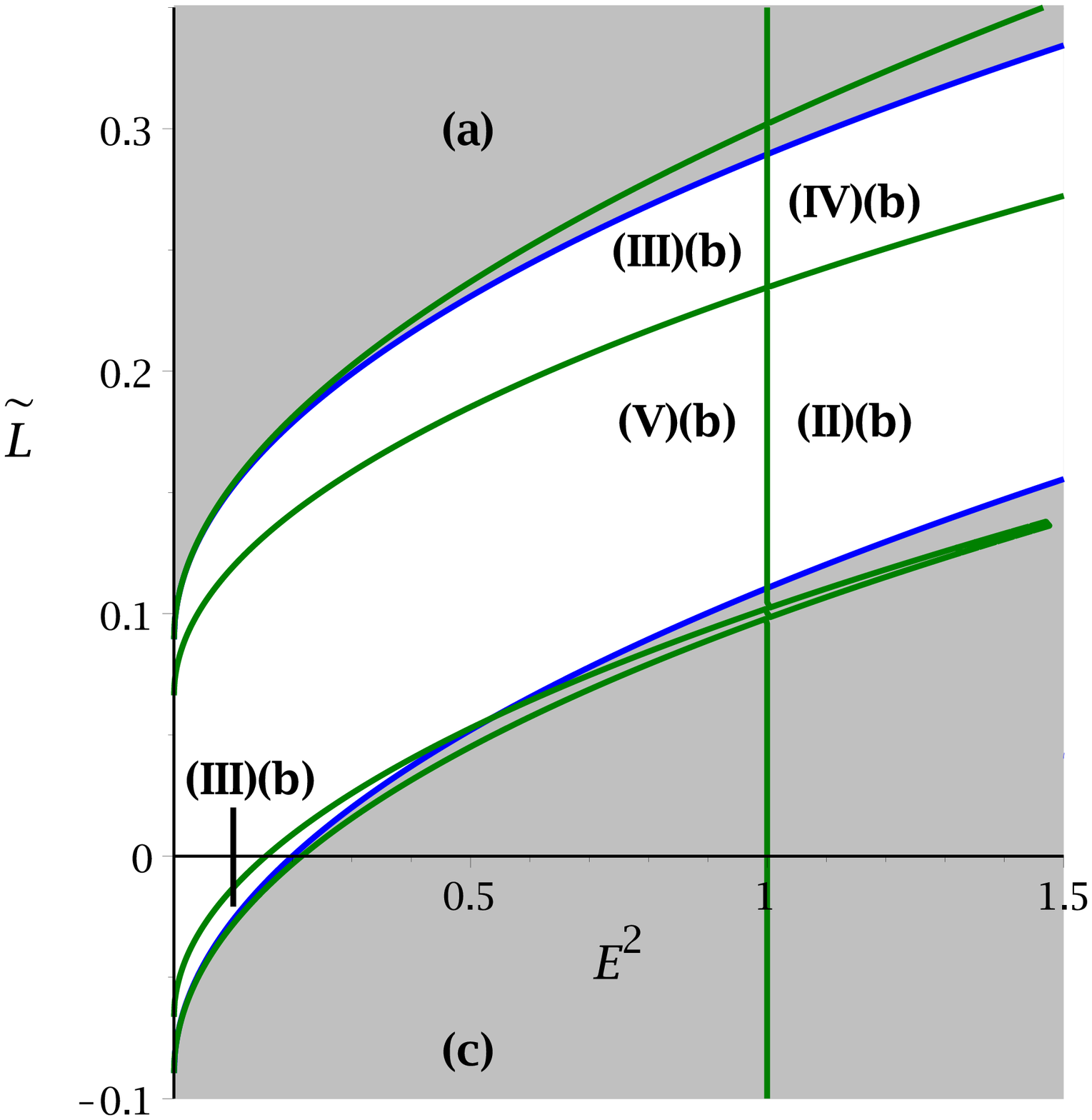}
	}
	\subfigure[$\delta=1$, $\ta=0.078$, $\tK=0.008$ , ${\tb=-0.08}$]{
		\includegraphics[width=0.31\textwidth]{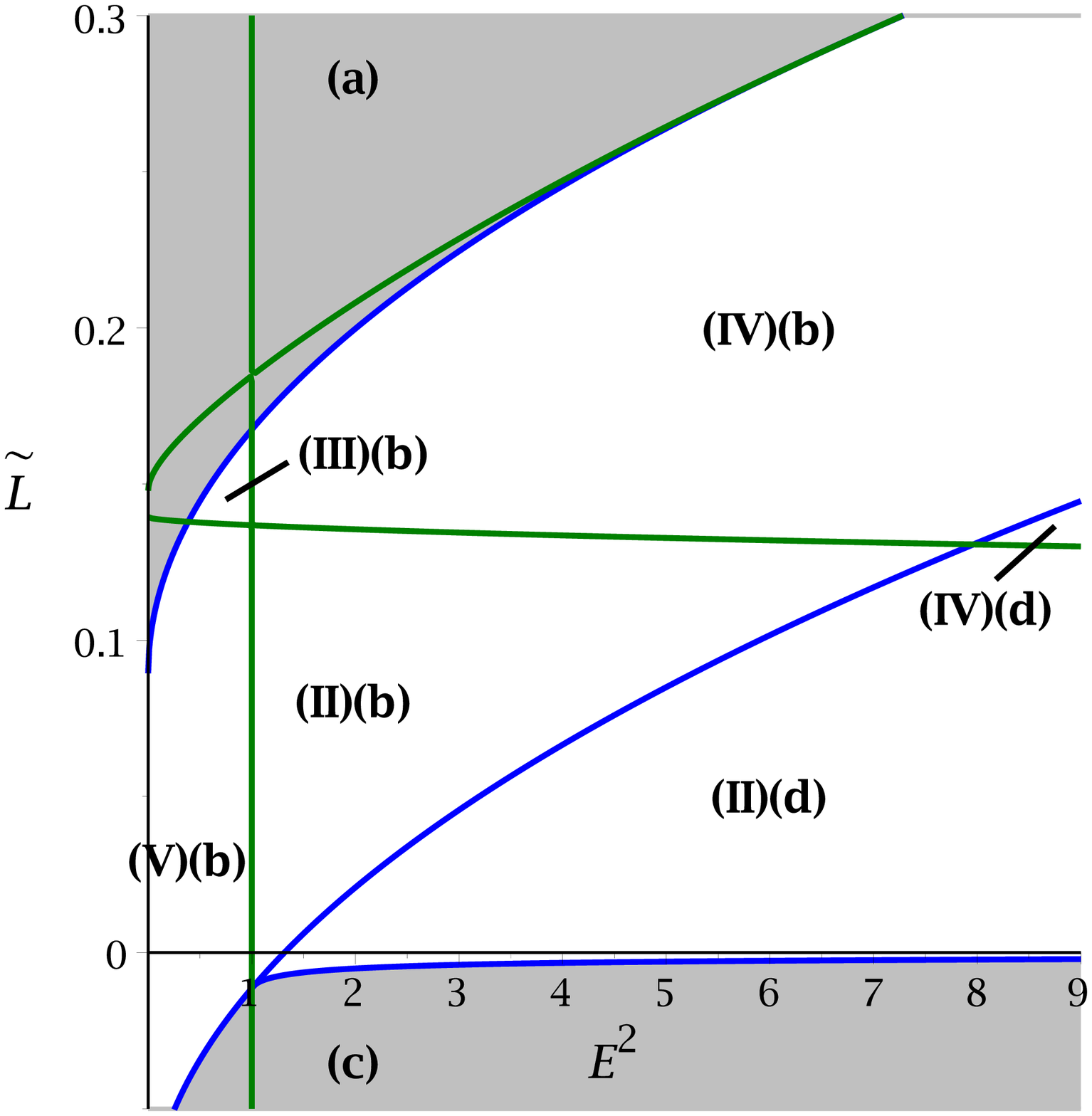}
	}
	\caption{Combined parametric $\tL$-$E^2$-diagrams of the $\tr$-motion (green lines) and $\vartheta$-motion (blue lines). In region (I) the polynomial $R$ has no real zeros. Two zeros are possible in the regions (II) and (V).  In the regions (II) and (V) $R$ has two zeros. The $\vartheta$-equation does not allow geodesic motion inside the grey areas (a) and (c). The orbits cross $\vartheta=\frac{\pi}{2}$, but not $\tr=0$ in regions marked with (b). In contrast, in regions marked with (d) $\tr=0$ can be crossed but $\vartheta=\frac{\pi}{2}$ is never crossed.}
 \label{pic:parametric-diagrams}
\end{figure}

The type of an orbit depends on how many real zeros the polynomial $R$ has. The number of zeros changes if double zeros appear, that is
\begin{equation}
	R(\tr)=0 \quad \text{and} \quad \frac{\dd R(\tr)}{\dd\tr}=0 \, .
	\label{eqn:doublezero}
\end{equation}
In the parametric $\tL$-$E^2$-diagrams based on \eqref{eqn:doublezero} five regions with different numbers of zeros can be distinguished. In region (I), the polynomial $R$ has no zeros. Region (II) has one positive and one negative zero, whereas region (V) has two positive zeros. In the remaining regions (III) and (IV) four real zeros are possible. Depending on the parameters of the metric and the test particle, region (III) has either four positive or two positive and two negative zeros. Region (IV) allows the following constellations: one negative and three positive zeros, three negative and one positive zero, or even four negative zeros. Figure \ref{pic:parametric-diagrams} shows three examples of the  parametric $\tL$-$E^2$-diagrams of the $\tr$-motion combined with the parametric diagrams of the $\vartheta$-motion.\\

Furthermore, the possible orbits can be visualized by an effective potential, which consists of two parts $V_+$ and $V_-$, determined by the equation
\begin{equation}
	R(\tr)=\left(\tr^2-2\tb\tr+\ta^2 \right) ^2 \left(E-V_+\right) \left(E-V_-\right) \, ,
\end{equation}
and thus
\begin{equation}
	V_\pm = \frac{ \ta\tL \pm \sqrt{ \left( \tr^2-\left(1+2\tb\tr\right)+\ta^2 \right)  \left( \delta\tr^2-2\delta\tb\tr +\tK \right) }}{\tr^2-2\tb\tr+\ta^2} \, .
\end{equation}
The area in between the two parts of the potential has $R(\tr)<0$, so there geodesic motion is not possible. In the limit $\tr \rightarrow \pm \infty$, $V_+$ converges to $\sqrt{\delta}$ and $V_-$ converges to $-\sqrt{\delta}$. If the parameter $a$ or $L$ changes its sign, the effective potential is mirrored at the $\tr$-axis. $V^+$ and $V^-$ meet at the horizons $\tr_\pm$ and if ${\tK \leq \delta\tb^2}$ they also meet at 
\begin{equation}
	\tr_{a,b}=\tb\pm\sqrt{\tb^2-\frac{\tK}{\delta}}
\end{equation}
leaving a gap between $\tr_a$ and $\tr_b$. Another peculiarity of the EMDA spacetime is the divergencies in the effective potential for ${a^2\leq\tb^2}$ at
\begin{equation}
	\tr_{c,d}=\tb\pm\sqrt{\tb^2-a^2} \, .
\end{equation}

\begin{figure}[h]
	\centering
	\subfigure[$\delta=1$, $\ta=0.4$, $\tb=-0.08$, $\tK=1$, $\tL=0.75$]{
		\includegraphics[width=0.31\textwidth]{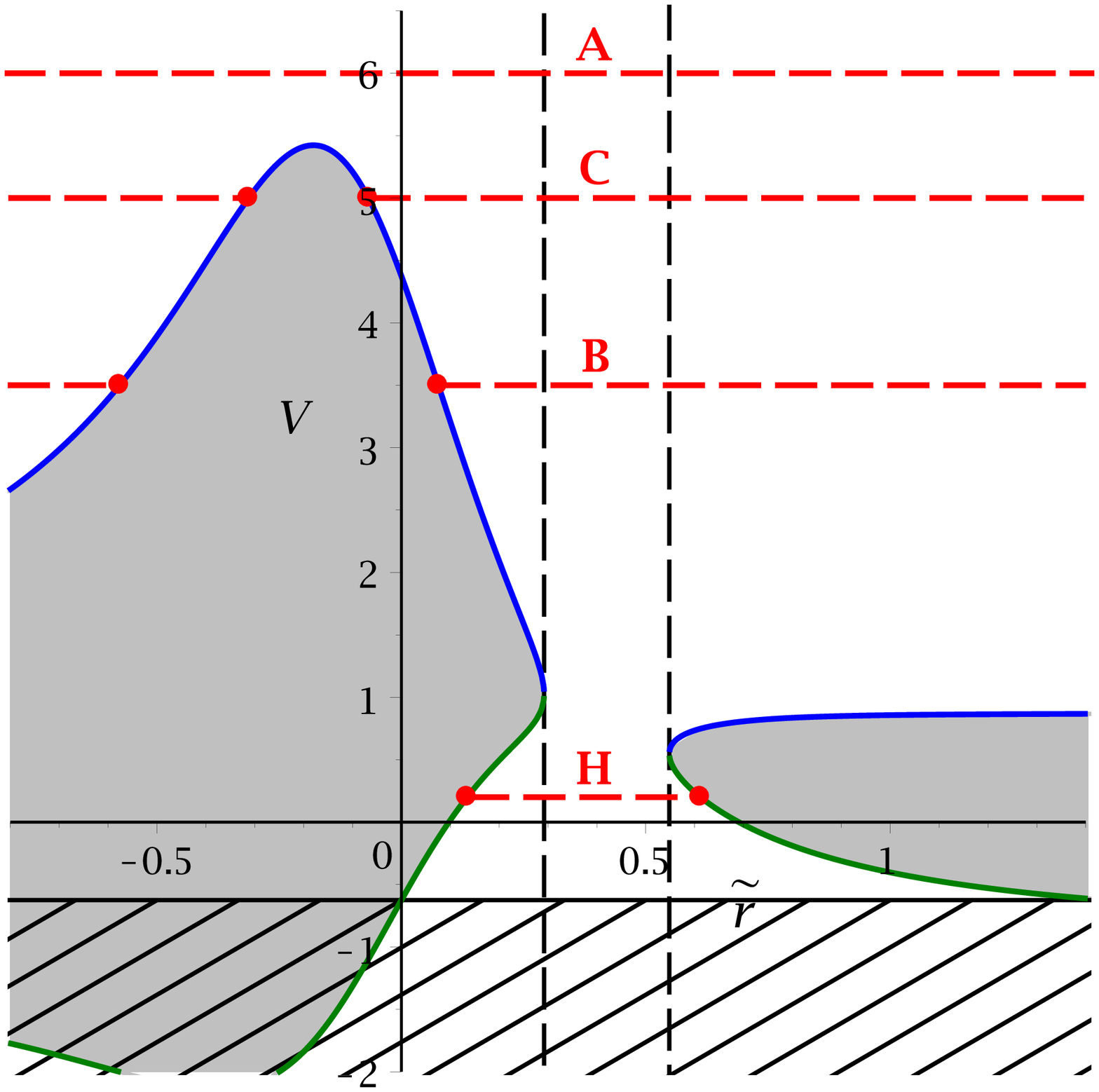}
	}
	\subfigure[$\delta=1$, $\ta=0.4$, $\tb=-0.05$, $\tK=2$, $\tL=1.5$]{
		\includegraphics[width=0.31\textwidth]{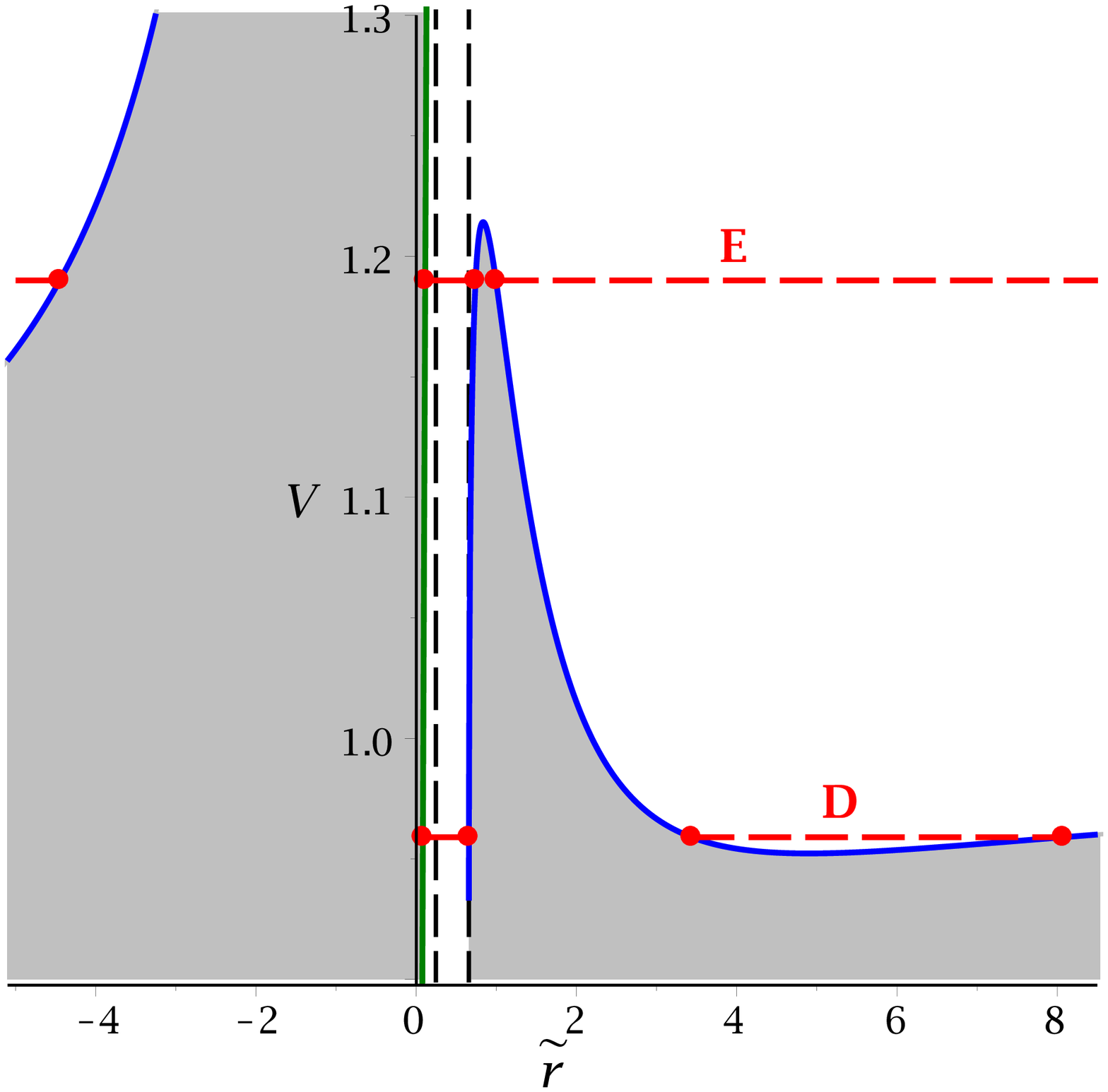}
	}
	\subfigure[$\delta=1$, $\ta=0.4$, $\tb=-0.08$, $\tK=2$, $\tL=2.9$]{
		\includegraphics[width=0.31\textwidth]{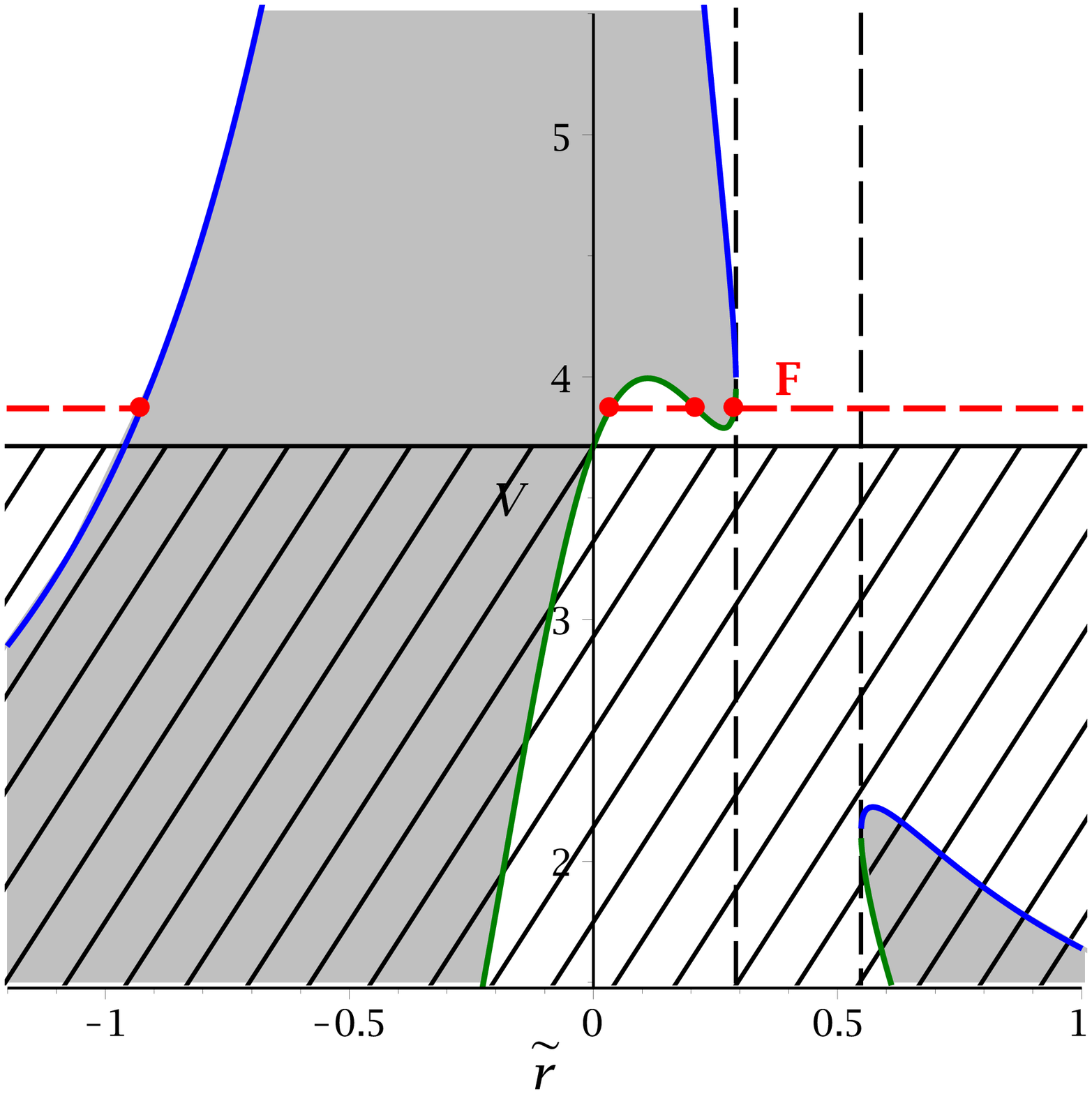}
	}

	\subfigure[$\delta=1$, $\ta=0.4$, $\tb=-0.08$, ${\tK=0.006}$, $\tL=0.1$]{
		\includegraphics[width=0.31\textwidth]{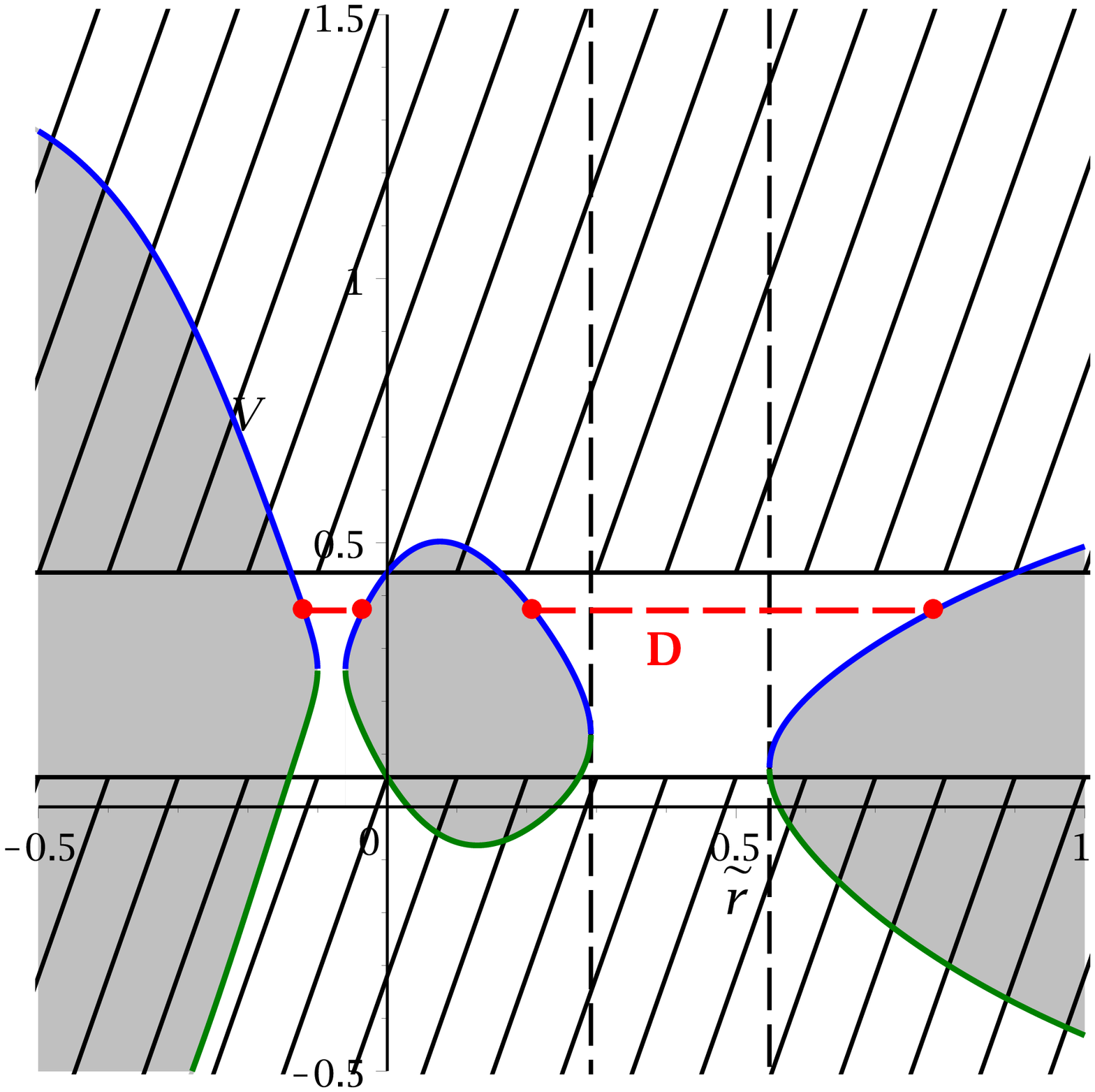}
	}
	\subfigure[$\delta=1$, $\ta=0.078$, $\tb=-0.08$, ${\tK=0.008}$, $\tL=0.1$]{
		\includegraphics[width=0.31\textwidth]{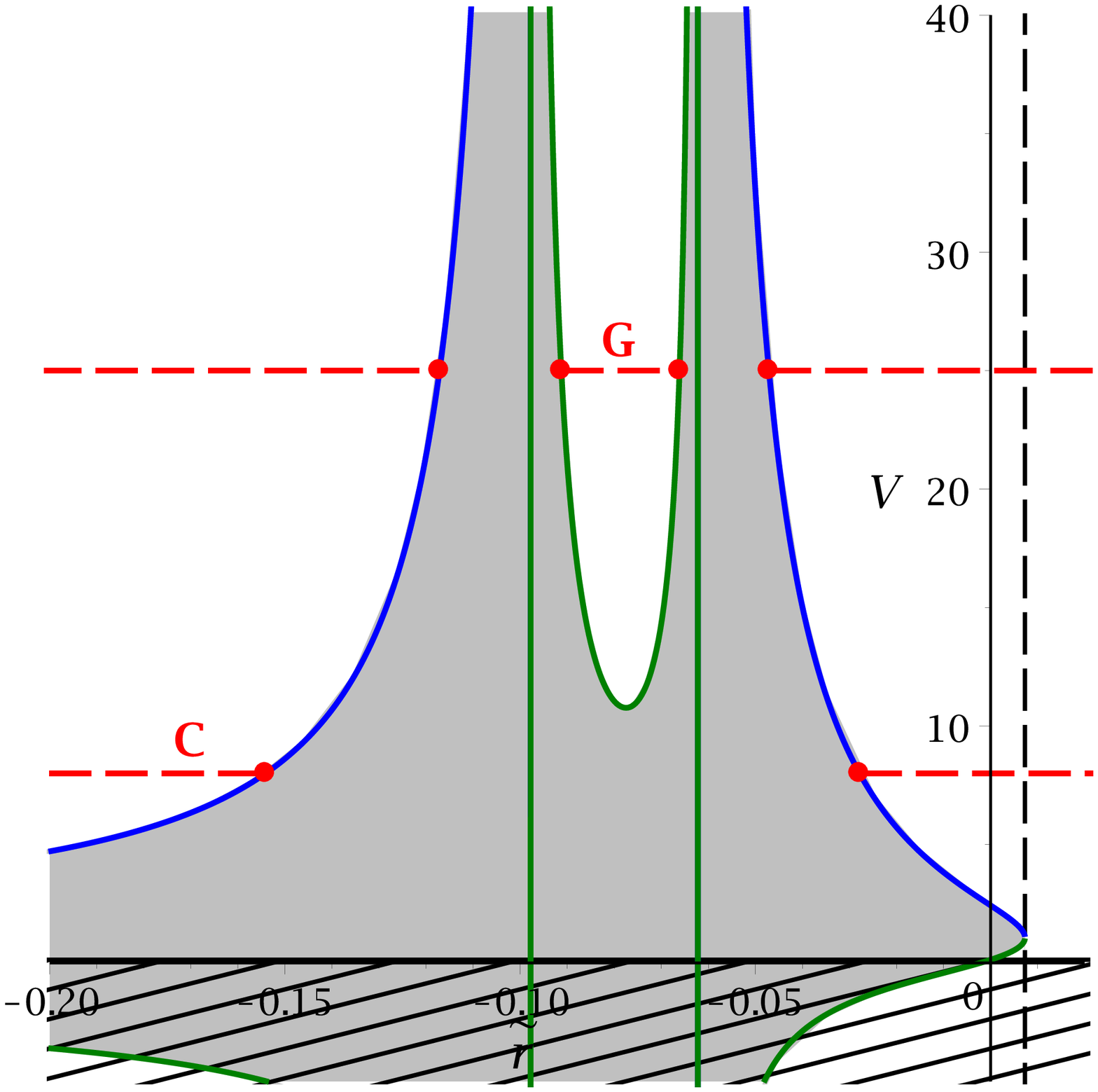}
	}
	\subfigure[$\delta=1$, $\ta=0.078$, $\tb=-0.08$, ${\tK=0.006}$, $\tL=0.1$]{
		\includegraphics[width=0.31\textwidth]{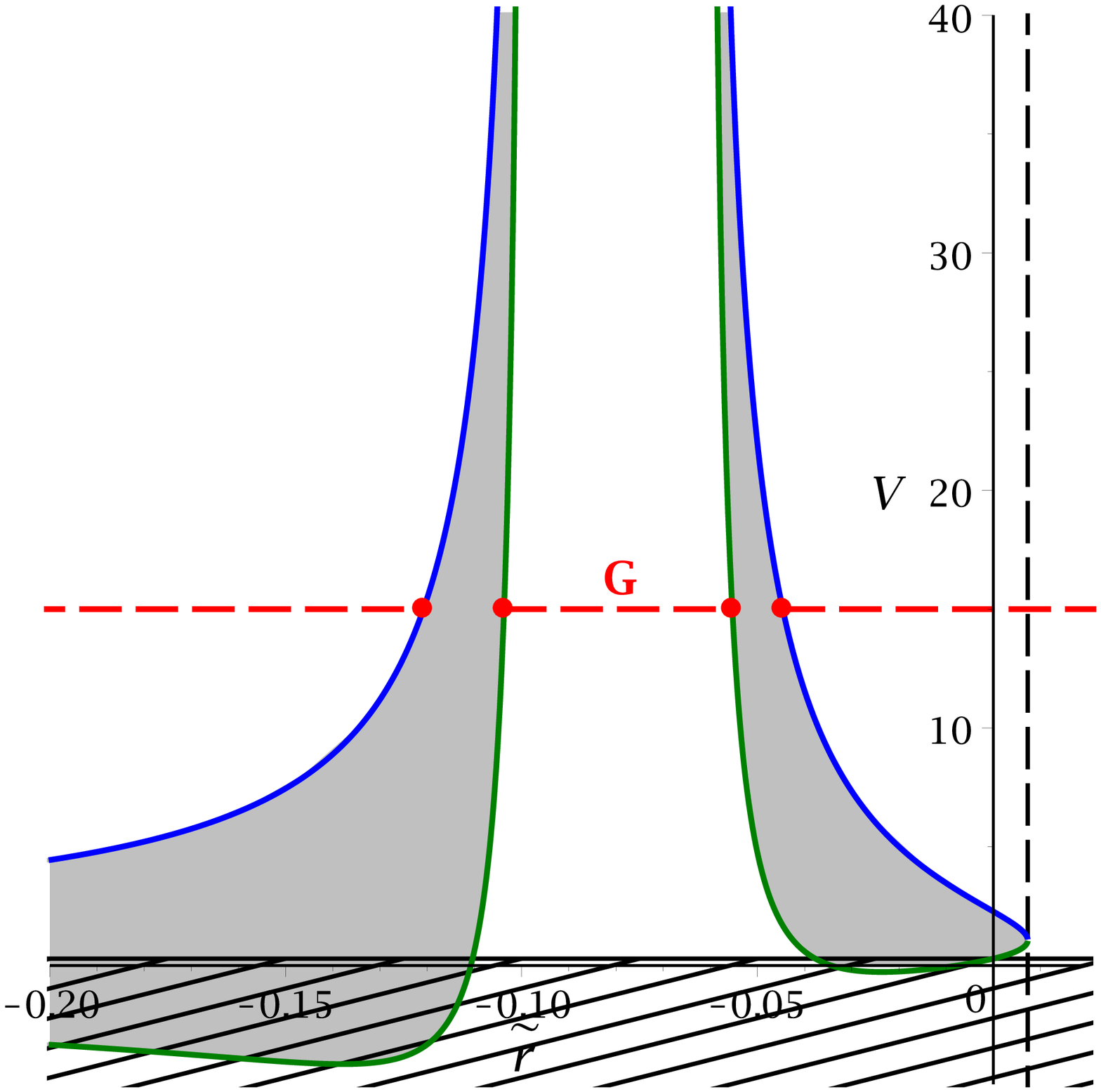}
	}
	\caption{Plots of the effective potentials with examples of energies for the different orbit types of table \ref{tab:orbit-types}. The blue curves correspond to $V_+(\tr)$ and the green curves correspond to $V_-(\tr)$. The grey area between $V_+$ and $V_-$ marks a zone, where geodesic motion is not possible since there $R(\tr) <0$. In the dashed area, the motion is forbidden by the $\vartheta$-equation ($\Theta(\vartheta) <0$). The red dashed lines represent the different energies. The red dots are the turning points of the orbits and accordingly the zeros of the polynomial $R$.}
 \label{pic:potentials}
\end{figure}

Taking all the information into consideration, we can now identify all possible orbits in each region of the parametric $\tL$-$E^2$-diagrams (below it is assumed that $\tr_i<\tr_{i+1}$):
\begin{enumerate}
	\item Region (I)(d): There are no real zeros and $R(\tr)>0$ for all $\tr$. The only possible orbit is a transit orbit which crosses $\tr=0$, but not $\vartheta=\frac{\pi}{2}$.
	\item Region (II)(b): $R(\tr)$ has a negative zero $\tr_1$ and a positive zero $\tr_2$, and $R(\tr)\geq0$ for $\tr\in(-\infty,\tr_1]$ and $\tr\in[\tr_2,\infty)$. The geodesics cross the equatorial plane. For $\tr<0$ an EO is possible and for $\tr\geq0$ a TEO is possible, whose  turning point can coincide with the inner horizon $\tr_-$. In the special case $\tr_2=0$, the TEO turns into a TO and the motion is confined to the equatorial plane.
	\item Region (II)(d): $R(\tr)$ has two negative zeros $\tr_1$, $\tr_2$ with $R(\tr)\geq0$ for $\tr\in(-\infty,\tr_1]$ and $\tr\in[\tr_2,\infty)$. There is an EO and a CTEO; this time the equatorial plane is not crossed.
	\item Region (III)(b): $R$ has four zeros $\tr_1$, $\tr_2$, $\tr_3$, $\tr_4$ and $R(\tr)\geq0$ for $\tr\in [\tr_1, \tr_2]$ and $\tr\in [\tr_3, \tr_4]$. If all four zeros are positive, there is a MBO between $r_1$ and $r_2$  (one of the zeros can coincide with a horizon), and a BO between $\tr_3$ and $\tr_4$. It is also possible for $\tb\neq0$, that $\tr_1$, $\tr_2$ are negative and $\tr_3$, $\tr_4$ are positive. Then there is a BO between $\tr_1$ and $\tr_2$, and an MBO between $\tr_3$ and $\tr_4$ (one of the zeros can coincide with a horizon). In the special case $\tK=(\ta E-\tL)^2$, the BO turns into a TO since $\tr_2=0$. In this region the geodesics cross $\vartheta=\frac{\pi}{2}$.
	\item Region (IV)(b): $R$ has four zeros  $\tr_1$, $\tr_2$, $\tr_3$, $\tr_4$ and $R(\tr)\geq0$ for $\tr\in (-\infty, \tr_1]$, $\tr\in [\tr_2, \tr_3]$ and $\tr\in [r_4, \infty)$. Either one zero is negative and three zeros are positive, or three zeros are negative and one zero is positive. The latter is only possible if $\tb\neq0$. EOs exist for $\tr\in (-\infty, \tr_1]$ (this one resides at negative $\tr$) and $\tr\in [\tr_4, \infty)$ if $\tr_4>\tr_+$. If $\tr_4\leq\tr_+$, then there is a TEO. Depending on the position of the zeros, various orbits are possible between $\tr_2$ and $\tr_3$. There can be MBOs  (one of the zeros can coincide with a horizon), BOs are hidden behind the horizons with $\tr_2,\tr_3 <\tr_-$ and even TOs are present if  $\tK=(\ta E-\tL)^2$. For $\tb\neq0$  it is possible that  $\tr_2,\tr_3 <0$ and therefore a BO for negative $\tr$ exists. In this region the geodesics cross $\vartheta=\frac{\pi}{2}$.
	\item Region (IV)(d): This regions only occurs if $\tb\neq0$; here $R$ has four negative zeros  $\tr_1$, $\tr_2$, $\tr_3$, $\tr_4$.  $R(\tr)\geq0$ for $\tr\in (-\infty, \tr_1]$, $\tr\in [\tr_2, \tr_3]$ and $\tr\in [r_4, \infty)$. An EO, a BO and a CTEO exist. Here $\tr=0$ can be crossed, but not $\vartheta=\frac{\pi}{2}$.
	\item Region (V)(b): $R$ has two positive zeros $\tr_1$, $\tr_2$ and $R(\tr)\geq0$ for $\tr\in [\tr_1, \tr_2]$. The possible orbit is a MBO, where one or both ($\tL=0$) zeros can coincide with the horizons. If for  $\tK=(\ta E-\tL)^2$ we have $\tr_1=0$, the orbit turns into a TO. The geodesics cross the equatorial plane.
\end{enumerate}
Table \ref{tab:orbit-types} shows an overview of all possible orbit types in the EMDA spacetime. In figure \ref{pic:potentials} examples for the corresponding energy for each orbit type are plotted together with the effective potentials. Note that the repulsive nature of the singularity at negative $\tr$ can be observed in figure \ref{pic:potentials}(e) and \ref{pic:potentials}(f), where potential barriers prevent the test particles from falling into the singularity. The corresponding plot of the singularity, which in this case consists of two closed surfaces, is shown in figure \ref{pic:singularity-shape}(c). The energy chosen for the orbit type G allows an EO at negative $\tr$, a BO at negative $\tr$ enclosed by the surfaces of singularity and a CTEO.

\begin{table}[h!]
\begin{center}
\begin{tabular}{|lcccll|}\hline
type & $\delta$ & zeros & region  & range of $\tr$ & orbit \\
\hline\hline
A & 1, 0 & 0 & Id &
\begin{pspicture}(-4,-0.2)(3.5,0.2)
\psline[linewidth=0.5pt]{->}(-4,0)(3.5,0)
\psline[linewidth=0.5pt](-2,-0.2)(-2,0.2)
\psline[linewidth=0.5pt,doubleline=true](0.3,-0.2)(0.3,0.2)
\psline[linewidth=0.5pt,doubleline=true](1.2,-0.2)(1.2,0.2)
\psline[linewidth=1.2pt]{-}(-4,0)(3.5,0)
\end{pspicture}
  & TrO
\\  \hline
B  & 1, 0 & 2 & IIb &
\begin{pspicture}(-4,-0.2)(3.5,0.2)
\psline[linewidth=0.5pt]{->}(-4,0)(3.5,0)
\psline[linewidth=0.5pt](-2,-0.2)(-2,0.2)
\psline[linewidth=0.5pt,doubleline=true](0.3,-0.2)(0.3,0.2)
\psline[linewidth=0.5pt,doubleline=true](1.2,-0.2)(1.2,0.2)
\psline[linewidth=1.2pt]{-*}(-4,0)(-2.5,0)
\psline[linewidth=1.2pt]{*-}(0,0)(3.5,0)
\end{pspicture}
& EO, TEO 
\\ 
B$_-$ & 1, 0 & &  & 
\begin{pspicture}(-4,-0.2)(3.5,0.2)
\psline[linewidth=0.5pt]{->}(-4,0)(3.5,0)
\psline[linewidth=0.5pt](-2,-0.2)(-2,0.2)
\psline[linewidth=0.5pt,doubleline=true](0.3,-0.2)(0.3,0.2)
\psline[linewidth=0.5pt,doubleline=true](1.2,-0.2)(1.2,0.2)
\psline[linewidth=1.2pt]{-*}(-4,0)(-2.5,0)
\psline[linewidth=1.2pt]{*-}(0.3,0)(3.5,0)
\end{pspicture}
  & EO, TEO
\\ 
B$_0$ & 1, 0 & &  & 
\begin{pspicture}(-4,-0.2)(3.5,0.2)
\psline[linewidth=0.5pt]{->}(-4,0)(3.5,0)
\psline[linewidth=0.5pt](-2,-0.2)(-2,0.2)
\psline[linewidth=0.5pt,doubleline=true](0.3,-0.2)(0.3,0.2)
\psline[linewidth=0.5pt,doubleline=true](1.2,-0.2)(1.2,0.2)
\psline[linewidth=1.2pt]{-*}(-4,0)(-2.5,0)
\psline[linewidth=1.2pt]{*-}(-2,0)(3.5,0)
\end{pspicture}
  & EO, TO
\\ \hline
C & 1, 0 & 2 & IId &
\begin{pspicture}(-4,-0.2)(3.5,0.2)
\psline[linewidth=0.5pt]{->}(-4,0)(3.5,0)
\psline[linewidth=0.5pt](-2,-0.2)(-2,0.2)
\psline[linewidth=0.5pt,doubleline=true](0.3,-0.2)(0.3,0.2)
\psline[linewidth=0.5pt,doubleline=true](1.2,-0.2)(1.2,0.2)
\psline[linewidth=1.2pt]{-*}(-4,0)(-3.5,0)
\psline[linewidth=1.2pt]{*-}(-2.5,0)(3.5,0)
\end{pspicture}
  & EO, CTEO
\\ \hline
D & 1 & 4 & IIIb &
\begin{pspicture}(-4,-0.2)(3.5,0.2)
\psline[linewidth=0.5pt]{->}(-4,0)(3.5,0)
\psline[linewidth=0.5pt](-2,-0.2)(-2,0.2)
\psline[linewidth=0.5pt,doubleline=true](0.3,-0.2)(0.3,0.2)
\psline[linewidth=0.5pt,doubleline=true](1.2,-0.2)(1.2,0.2)
\psline[linewidth=1.2pt]{*-*}(0,0)(1.5,0)
\psline[linewidth=1.2pt]{*-*}(2,0)(3,0)
\end{pspicture}
& MBO, BO 
\\
D$_-$ & 1 &  &  &
\begin{pspicture}(-4,-0.2)(3.5,0.2)
\psline[linewidth=0.5pt]{->}(-4,0)(3.5,0)
\psline[linewidth=0.5pt](-2,-0.2)(-2,0.2)
\psline[linewidth=0.5pt,doubleline=true](0.3,-0.2)(0.3,0.2)
\psline[linewidth=0.5pt,doubleline=true](1.2,-0.2)(1.2,0.2)
\psline[linewidth=1.2pt]{*-*}(0.3,0)(1.5,0)
\psline[linewidth=1.2pt]{*-*}(2,0)(3,0)
\end{pspicture}
& MBO, BO 
\\
D$_+$ &  1 &  &  &
\begin{pspicture}(-4,-0.2)(3.5,0.2)
\psline[linewidth=0.5pt]{->}(-4,0)(3.5,0)
\psline[linewidth=0.5pt](-2,-0.2)(-2,0.2)
\psline[linewidth=0.5pt,doubleline=true](0.3,-0.2)(0.3,0.2)
\psline[linewidth=0.5pt,doubleline=true](1.2,-0.2)(1.2,0.2)
\psline[linewidth=1.2pt]{*-*}(0,0)(1.2,0)
\psline[linewidth=1.2pt]{*-*}(2,0)(3,0)
\end{pspicture}
& MBO, BO
\\
D & 1 & &  &
\begin{pspicture}(-4,-0.2)(3.5,0.2)
\psline[linewidth=0.5pt]{->}(-4,0)(3.5,0)
\psline[linewidth=0.5pt](-2,-0.2)(-2,0.2)
\psline[linewidth=0.5pt,doubleline=true](0.3,-0.2)(0.3,0.2)
\psline[linewidth=0.5pt,doubleline=true](1.2,-0.2)(1.2,0.2)
\psline[linewidth=1.2pt]{*-*}(-3.5,0)(-2.5,0)
\psline[linewidth=1.2pt]{*-*}(0,0)(1.5,0)
\end{pspicture}
  &  BO, MBO
\\
D$_0$ & 1 & &  &
\begin{pspicture}(-4,-0.2)(3.5,0.2)
\psline[linewidth=0.5pt]{->}(-4,0)(3.5,0)
\psline[linewidth=0.5pt](-2,-0.2)(-2,0.2)
\psline[linewidth=0.5pt,doubleline=true](0.3,-0.2)(0.3,0.2)
\psline[linewidth=0.5pt,doubleline=true](1.2,-0.2)(1.2,0.2)
\psline[linewidth=1.2pt]{*-*}(-3.5,0)(-2,0)
\psline[linewidth=1.2pt]{*-*}(0,0)(1.5,0)
\end{pspicture}
  &  TO, MBO
\\
D$_0$ & 1 & &  &
\begin{pspicture}(-4,-0.2)(3.5,0.2)
\psline[linewidth=0.5pt]{->}(-4,0)(3.5,0)
\psline[linewidth=0.5pt](-2,-0.2)(-2,0.2)
\psline[linewidth=0.5pt,doubleline=true](0.3,-0.2)(0.3,0.2)
\psline[linewidth=0.5pt,doubleline=true](1.2,-0.2)(1.2,0.2)
\psline[linewidth=1.2pt]{*-*}(-3.5,0)(-2.5,0)
\psline[linewidth=1.2pt]{*-*}(-2,0)(1.5,0)
\end{pspicture}
  &  BO, TO
\\
D$_-$ & 1 & &  &
\begin{pspicture}(-4,-0.2)(3.5,0.2)
\psline[linewidth=0.5pt]{->}(-4,0)(3.5,0)
\psline[linewidth=0.5pt](-2,-0.2)(-2,0.2)
\psline[linewidth=0.5pt,doubleline=true](0.3,-0.2)(0.3,0.2)
\psline[linewidth=0.5pt,doubleline=true](1.2,-0.2)(1.2,0.2)
\psline[linewidth=1.2pt]{*-*}(-3.5,0)(-2.5,0)
\psline[linewidth=1.2pt]{*-*}(0.3,0)(1.5,0)
\end{pspicture}
  &  BO, MBO
\\
D$_0-$ & 1 & &  &
\begin{pspicture}(-4,-0.2)(3.5,0.2)
\psline[linewidth=0.5pt]{->}(-4,0)(3.5,0)
\psline[linewidth=0.5pt](-2,-0.2)(-2,0.2)
\psline[linewidth=0.5pt,doubleline=true](0.3,-0.2)(0.3,0.2)
\psline[linewidth=0.5pt,doubleline=true](1.2,-0.2)(1.2,0.2)
\psline[linewidth=1.2pt]{*-*}(-3.5,0)(-2,0)
\psline[linewidth=1.2pt]{*-*}(0.3,0)(1.5,0)
\end{pspicture}
  &  TO, MBO
\\
D$_+$ & 1 & &  &
\begin{pspicture}(-4,-0.2)(3.5,0.2)
\psline[linewidth=0.5pt]{->}(-4,0)(3.5,0)
\psline[linewidth=0.5pt](-2,-0.2)(-2,0.2)
\psline[linewidth=0.5pt,doubleline=true](0.3,-0.2)(0.3,0.2)
\psline[linewidth=0.5pt,doubleline=true](1.2,-0.2)(1.2,0.2)
\psline[linewidth=1.2pt]{*-*}(-3.5,0)(-2.5,0)
\psline[linewidth=1.2pt]{*-*}(0,0)(1.2,0)
\end{pspicture}
  &  BO, MBO
\\
D$_0+$ & 1 & &  &
\begin{pspicture}(-4,-0.2)(3.5,0.2)
\psline[linewidth=0.5pt]{->}(-4,0)(3.5,0)
\psline[linewidth=0.5pt](-2,-0.2)(-2,0.2)
\psline[linewidth=0.5pt,doubleline=true](0.3,-0.2)(0.3,0.2)
\psline[linewidth=0.5pt,doubleline=true](1.2,-0.2)(1.2,0.2)
\psline[linewidth=1.2pt]{*-*}(-3.5,0)(-2,0)
\psline[linewidth=1.2pt]{*-*}(0,0)(1.2,0)
\end{pspicture}
  &  TO, MBO
\\ \hline
E & 1, 0 & 4 & IVb &
\begin{pspicture}(-4,-0.2)(3.5,0.2)
\psline[linewidth=0.5pt]{->}(-4,0)(3.5,0)
\psline[linewidth=0.5pt](-2,-0.2)(-2,0.2)
\psline[linewidth=0.5pt,doubleline=true](0.3,-0.2)(0.3,0.2)
\psline[linewidth=0.5pt,doubleline=true](1.2,-0.2)(1.2,0.2)
\psline[linewidth=1.2pt]{-*}(-4,0)(-2.5,0)
\psline[linewidth=1.2pt]{*-*}(0,0)(1.5,0)
\psline[linewidth=1.2pt]{*-}(2,0)(3.5,0)
\end{pspicture}
  & EO, MBO, EO
\\
E$_-$ & 1, 0 &  &  &
\begin{pspicture}(-4,-0.2)(3.5,0.2)
\psline[linewidth=0.5pt]{->}(-4,0)(3.5,0)
\psline[linewidth=0.5pt](-2,-0.2)(-2,0.2)
\psline[linewidth=0.5pt,doubleline=true](0.3,-0.2)(0.3,0.2)
\psline[linewidth=0.5pt,doubleline=true](1.2,-0.2)(1.2,0.2)
\psline[linewidth=1.2pt]{-*}(-4,0)(-2.5,0)
\psline[linewidth=1.2pt]{*-*}(0.3,0)(1.5,0)
\psline[linewidth=1.2pt]{*-}(2,0)(3.5,0)
\end{pspicture}
  & EO, MBO, EO
\\
E$_+$ & 1, 0 &  &  &
\begin{pspicture}(-4,-0.2)(3.5,0.2)
\psline[linewidth=0.5pt]{->}(-4,0)(3.5,0)
\psline[linewidth=0.5pt](-2,-0.2)(-2,0.2)
\psline[linewidth=0.5pt,doubleline=true](0.3,-0.2)(0.3,0.2)
\psline[linewidth=0.5pt,doubleline=true](1.2,-0.2)(1.2,0.2)
\psline[linewidth=1.2pt]{-*}(-4,0)(-2.5,0)
\psline[linewidth=1.2pt]{*-*}(0,0)(1.2,0)
\psline[linewidth=1.2pt]{*-}(2,0)(3.5,0)
\end{pspicture}
  & EO, MBO, EO
\\ \hline
F & 1, 0 & 4 & IVb &
\begin{pspicture}(-4,-0.2)(3.5,0.2)
\psline[linewidth=0.5pt]{->}(-4,0)(3.5,0)
\psline[linewidth=0.5pt](-2,-0.2)(-2,0.2)
\psline[linewidth=0.5pt,doubleline=true](0.3,-0.2)(0.3,0.2)
\psline[linewidth=0.5pt,doubleline=true](1.2,-0.2)(1.2,0.2)
\psline[linewidth=1.2pt]{-*}(-4,0)(-2.5,0)
\psline[linewidth=1.2pt]{*-*}(-1.5,0)(-0.5,0)
\psline[linewidth=1.2pt]{*-}(0,0)(3.5,0)
\end{pspicture}
  & EO, BO, TEO
\\
F$_-$ & 1, 0 &  &  &
\begin{pspicture}(-4,-0.2)(3.5,0.2)
\psline[linewidth=0.5pt]{->}(-4,0)(3.5,0)
\psline[linewidth=0.5pt](-2,-0.2)(-2,0.2)
\psline[linewidth=0.5pt,doubleline=true](0.3,-0.2)(0.3,0.2)
\psline[linewidth=0.5pt,doubleline=true](1.2,-0.2)(1.2,0.2)
\psline[linewidth=1.2pt]{-*}(-4,0)(-2.5,0)
\psline[linewidth=1.2pt]{*-*}(-1.5,0)(-0.5,0)
\psline[linewidth=1.2pt]{*-}(0.3,0)(3.5,0)
\end{pspicture}
  & EO, BO, TEO
\\
F$_0$ & 1, 0 &  &  &
\begin{pspicture}(-4,-0.2)(3.5,0.2)
\psline[linewidth=0.5pt]{->}(-4,0)(3.5,0)
\psline[linewidth=0.5pt](-2,-0.2)(-2,0.2)
\psline[linewidth=0.5pt,doubleline=true](0.3,-0.2)(0.3,0.2)
\psline[linewidth=0.5pt,doubleline=true](1.2,-0.2)(1.2,0.2)
\psline[linewidth=1.2pt]{-*}(-4,0)(-2.5,0)
\psline[linewidth=1.2pt]{*-*}(-2,0)(-0.5,0)
\psline[linewidth=1.2pt]{*-}(0,0)(3.5,0)
\end{pspicture}
  & EO, TO, TEO
\\
F$_{0-}$ & 1, 0 &  &  &
\begin{pspicture}(-4,-0.2)(3.5,0.2)
\psline[linewidth=0.5pt]{->}(-4,0)(3.5,0)
\psline[linewidth=0.5pt](-2,-0.2)(-2,0.2)
\psline[linewidth=0.5pt,doubleline=true](0.3,-0.2)(0.3,0.2)
\psline[linewidth=0.5pt,doubleline=true](1.2,-0.2)(1.2,0.2)
\psline[linewidth=1.2pt]{-*}(-4,0)(-2.5,0)
\psline[linewidth=1.2pt]{*-*}(-2,0)(-0.5,0)
\psline[linewidth=1.2pt]{*-}(0.3,0)(3.5,0)
\end{pspicture}
  & EO, TO, TEO
\\
F & 1 &  &  &
\begin{pspicture}(-4,-0.2)(3.5,0.2)
\psline[linewidth=0.5pt]{->}(-4,0)(3.5,0)
\psline[linewidth=0.5pt](-0.5,-0.2)(-0.5,0.2)
\psline[linewidth=0.5pt,doubleline=true](0.3,-0.2)(0.3,0.2)
\psline[linewidth=0.5pt,doubleline=true](1.2,-0.2)(1.2,0.2)
\psline[linewidth=1.2pt]{-*}(-4,0)(-2.5,0)
\psline[linewidth=1.2pt]{*-*}(-2,0)(-1,0)
\psline[linewidth=1.2pt]{*-}(0,0)(3.5,0)
\end{pspicture}
  & EO, BO, TEO
\\
F$_{-}$ & 1 & &  &
\begin{pspicture}(-4,-0.2)(3.5,0.2)
\psline[linewidth=0.5pt]{->}(-4,0)(3.5,0)
\psline[linewidth=0.5pt](-0.5,-0.2)(-0.5,0.2)
\psline[linewidth=0.5pt,doubleline=true](0.3,-0.2)(0.3,0.2)
\psline[linewidth=0.5pt,doubleline=true](1.2,-0.2)(1.2,0.2)
\psline[linewidth=1.2pt]{-*}(-4,0)(-2.5,0)
\psline[linewidth=1.2pt]{*-*}(-2,0)(-1,0)
\psline[linewidth=1.2pt]{*-}(0.3,0)(3.5,0)
\end{pspicture}
  & EO, BO, TEO
\\
F$_{0}$ & 1 & &  &
\begin{pspicture}(-4,-0.2)(3.5,0.2)
\psline[linewidth=0.5pt]{->}(-4,0)(3.5,0)
\psline[linewidth=0.5pt](-0.5,-0.2)(-0.5,0.2)
\psline[linewidth=0.5pt,doubleline=true](0.3,-0.2)(0.3,0.2)
\psline[linewidth=0.5pt,doubleline=true](1.2,-0.2)(1.2,0.2)
\psline[linewidth=1.2pt]{-*}(-4,0)(-2.5,0)
\psline[linewidth=1.2pt]{*-*}(-2,0)(-1,0)
\psline[linewidth=1.2pt]{*-}(-0.5,0)(3.5,0)
\end{pspicture}
  & EO, BO, TO
\\
F$_{0-}$ & 1 & &  &
\begin{pspicture}(-4,-0.2)(3.5,0.2)
\psline[linewidth=0.5pt]{->}(-4,0)(3.5,0)
\psline[linewidth=0.5pt](-0.5,-0.2)(-0.5,0.2)
\psline[linewidth=0.5pt,doubleline=true](0.3,-0.2)(0.3,0.2)
\psline[linewidth=0.5pt,doubleline=true](1.2,-0.2)(1.2,0.2)
\psline[linewidth=1.2pt]{-*}(-4,0)(-2.5,0)
\psline[linewidth=1.2pt]{*-*}(-2,0)(-0.5,0)
\psline[linewidth=1.2pt]{*-}(0.3,0)(3.5,0)
\end{pspicture}
  & EO, TO, TEO
\\ \hline
G & 1 & 4 & IVd &
\begin{pspicture}(-4,-0.2)(3.5,0.2)
\psline[linewidth=0.5pt]{->}(-4,0)(3.5,0)
\psline[linewidth=0.5pt](-0.5,-0.2)(-0.5,0.2)
\psline[linewidth=0.5pt,doubleline=true](0.3,-0.2)(0.3,0.2)
\psline[linewidth=0.5pt,doubleline=true](1.2,-0.2)(1.2,0.2)
\psline[linewidth=1.2pt]{-*}(-4,0)(-3,0)
\psline[linewidth=1.2pt]{*-*}(-2.5,0)(-1.5,0)
\psline[linewidth=1.2pt]{*-}(-1,0)(3.5,0)
\end{pspicture}
  & EO, BO, CTEO
\\ \hline
H & 1, 0 & 2 & Vb &
\begin{pspicture}(-4,-0.2)(3.5,0.2)
\psline[linewidth=0.5pt]{->}(-4,0)(3.5,0)
\psline[linewidth=0.5pt](-2,-0.2)(-2,0.2)
\psline[linewidth=0.5pt,doubleline=true](0.3,-0.2)(0.3,0.2)
\psline[linewidth=0.5pt,doubleline=true](1.2,-0.2)(1.2,0.2)
\psline[linewidth=1.2pt]{*-*}(0,0)(1.5,0)
\end{pspicture}
& MBO 
\\
H$_\pm$ & 1, 0 &  &  &
\begin{pspicture}(-4,-0.2)(3.5,0.2)
\psline[linewidth=0.5pt]{->}(-4,0)(3.5,0)
\psline[linewidth=0.5pt](-2,-0.2)(-2,0.2)
\psline[linewidth=0.5pt,doubleline=true](0.3,-0.2)(0.3,0.2)
\psline[linewidth=0.5pt,doubleline=true](1.2,-0.2)(1.2,0.2)
\psline[linewidth=1.2pt]{*-*}(0.3,0)(1.2,0)
\end{pspicture}
& MBO
\\
H$_-$ & 1 &  &  &
\begin{pspicture}(-4,-0.2)(3.5,0.2)
\psline[linewidth=0.5pt]{->}(-4,0)(3.5,0)
\psline[linewidth=0.5pt](-2,-0.2)(-2,0.2)
\psline[linewidth=0.5pt,doubleline=true](0.3,-0.2)(0.3,0.2)
\psline[linewidth=0.5pt,doubleline=true](1.2,-0.2)(1.2,0.2)
\psline[linewidth=1.2pt]{*-*}(0.3,0)(1.5,0)
\end{pspicture}
& MBO
\\
H$_+$ & 1 &  &  &
\begin{pspicture}(-4,-0.2)(3.5,0.2)
\psline[linewidth=0.5pt]{->}(-4,0)(3.5,0)
\psline[linewidth=0.5pt](-2,-0.2)(-2,0.2)
\psline[linewidth=0.5pt,doubleline=true](0.3,-0.2)(0.3,0.2)
\psline[linewidth=0.5pt,doubleline=true](1.2,-0.2)(1.2,0.2)
\psline[linewidth=1.2pt]{*-*}(0,0)(1.2,0)
\end{pspicture}
& MBO
\\
H$_0$& 1 &  &  &
\begin{pspicture}(-4,-0.2)(3.5,0.2)
\psline[linewidth=0.5pt]{->}(-4,0)(3.5,0)
\psline[linewidth=0.5pt](-2,-0.2)(-2,0.2)
\psline[linewidth=0.5pt,doubleline=true](0.3,-0.2)(0.3,0.2)
\psline[linewidth=0.5pt,doubleline=true](1.2,-0.2)(1.2,0.2)
\psline[linewidth=1.2pt]{*-*}(-2,0)(1.5,0)
\end{pspicture}
& TO
\\
H$_{0+}$& 1 &  &  &
\begin{pspicture}(-4,-0.2)(3.5,0.2)
\psline[linewidth=0.5pt]{->}(-4,0)(3.5,0)
\psline[linewidth=0.5pt](-2,-0.2)(-2,0.2)
\psline[linewidth=0.5pt,doubleline=true](0.3,-0.2)(0.3,0.2)
\psline[linewidth=0.5pt,doubleline=true](1.2,-0.2)(1.2,0.2)
\psline[linewidth=1.2pt]{*-*}(-2,0)(1.2,0)
\end{pspicture}
& TO
\\ \hline\hline
\end{tabular}
\caption{Types of orbits of light and particles in the EMDA spacetime. The range of the orbits is represented by thick lines. The turning points are marked by thick dots. The two vertical double lines indicate the position of the horizons and the single vertical line corresponds to $\tr=0$. The column $\delta$ shows whether the orbit is possible for particles (1) and/or light (0).}
\label{tab:orbit-types}
\end{center}
\end{table}

\section{Solution of the geodesic equation}

In this section the analytical solutions of the geodesic equations \eqref{eqn:r-equation}-\eqref{eqn:t-equation} are given in terms of the Weierstra{\ss} $\wp$-, $\sigma$- and $\zeta$-functions.

\subsection{The $\tr$-equation}

The fourth order polynomial $R=\sum _{i=1}^4 a_i\tr^i$ on the right-hand side of equation \eqref{eqn:r-equation} with the coefficients
\begin{equation}
	\begin{split}
		a_4 &=E^2-\delta\\
		a_3 &=-4\tb \left(E^2-\delta\right)+\delta\\
		a_2 &=2\ta E \left(\ta E-\tL\right) - \tK-\ta^2\delta -2\delta\left(1+2\tb\right)+4\beta^2E^2\\
		a_1 &=\tK\left(1+2\tb\right)+2\ta^2\delta\tb -4\ta\tb E \left(\ta E-\tL\right)\\
		a_0 &= \ta^2\left[\left(\ta E-\tL\right)^2-\tK\right]
	\end{split}
\end{equation}
can be reduced to order three by substituting $\tr=\pm\frac{1}{x}+\tr_R$ (where $\tr_R$ is a zero of $R$): $R'= \sum _{i=0}^3 b_i x^i$. Then  $x=\frac{1}{b_3}\left( 4y-\frac{b_2}{3}\right)$ converts $R'$ into the Weierstra{\ss} form $P_3^{\tr}$ and \eqref{eqn:r-equation} acquires the form
\begin{equation}
	\left(\frac{dy}{d\gamma}\right)^2=4y^3-g_2^{\tr}y-g_3^{\tr}= P_3^{\tr} (y) \, ,
	\label{eqn:weierstrass-form}
\end{equation}
with
\begin{equation}
	g_2^{\tr}=\frac{b_2^2}{12} - \frac{b_1b_3}{4} \, , \qquad  g_3^{\tr}=\frac{b_1b_2b_3}{48} - \frac{b_0b_3^2}{16}-\frac{b_2^3}{216} \ .
\end{equation}
The elliptic differential equation \eqref{eqn:weierstrass-form} is solved by the Weierstra{\ss} $\wp$-function \cite{Markushevich:1967}
\begin{equation}
	y(\gamma) = \wp\left(\gamma - \gamma'_{\rm in}; g_2^{\tr}, g_3^{\tr}\right) \ ,
\end{equation}
where $\gamma'_{\rm in}=\gamma_{\rm in}+\int^\infty_{y_{\rm in}}{\frac{dy}{\sqrt{4y^3-g_2^{\tr}y-g_3^{\tr}}}}$ with $y_{\rm in}=\pm\frac{b_3}{4\tr_{\rm in}} + \frac{b_2}{12}$. Finally, resubstitution yields the solution of the $\tr$-equation \eqref{eqn:r-equation}
\begin{equation}
	\tr=\pm \frac{b_3}{4 \wp\left(\gamma - \gamma'_{\rm in}; g_2^{\tr}, g_3^{\tr}\right) - \frac{b_2}{3}} +\tr_R\ .
\end{equation}

\subsection{The $\vartheta$-equation}

First we substitute $\nu= \cos^2\!\vartheta$ to simplify the $\vartheta$-equation \eqref{eqn:theta-equation}
\begin{equation}
	\left( \frac{\dd\nu}{\dd\gamma} \right)^2 = 4\nu\left(1-\nu\right)\left(\tK-\delta\ta^2\nu\right)-4\nu\left[\tL - \ta E\left(1-\nu\right) \right]^2\, .
	\label{eqn:theta-polynomial}
\end{equation}
This transforms the right-hand side of the $\vartheta$-equation into a third order polynomial $\sum^3_{i=1}c_i\nu^i$ with the coefficients
\begin{eqnarray}
	c_3 &=& -4\ta^2\left( E^2-\delta \right)\\
	c_2 &=& -4\left[ 2\ta E\left( \tL - \ta E \right) + \ta^2 \delta +\tK\right]\\
	c_1 &=& 4\left[\tK-\left( \tL - \ta E\right)^2\right] \, .
\end{eqnarray}
The next step is to substitute  $\nu=\frac{1}{c_3}\left(4u-\frac{c_2}{3}\right)$ so that equation \eqref{eqn:theta-polynomial} acquires the standard Weierstra{\ss} form
\begin{equation}
	\left( \frac{\dd u}{\dd\gamma} \right)^2 =4u^3-g_2^\vartheta u -g_3^\vartheta = P_3^{\vartheta} (u) \, ,
	\label{eqn:theta-weierstrass}
\end{equation}
with
\begin{equation}
	g_2^\vartheta=\frac{c_2^2}{12} - \frac{c_1c_3}{4} \, , \qquad  g_3^\vartheta=\frac{c_1c_2c_3}{48}-\frac{c_2^3}{216} \ .
\end{equation}
Since equation \eqref{eqn:theta-weierstrass} is solved by the Weierstra{\ss} $\wp$-function, the solution $\vartheta(\gamma)$ of equation \eqref{eqn:theta-equation} is:
\begin{equation}
	\vartheta(\gamma) = \arccos\left( \pm \sqrt{\frac{1}{c_3}\left( 4\wp(\gamma-\gamma''_{\rm in};g_2^\vartheta,g_3^\vartheta) -\frac{c_2}{3} \right)} \right)
\end{equation}
where $\gamma''_{\rm in}=\gamma_{\rm in}+\int^\infty_{u_{\rm in}}{\frac{du}{\sqrt{4u^3-g_2^\vartheta y-g_3^\vartheta}}}$ and $u_{\rm in}=\frac{c_3}{4}\cos^2\vartheta_{\rm in} + \frac{c_2}{12}$.

\subsection{The $\varphi$-equation}
\label{sec:phisol}

The $\varphi$-equation \eqref{eqn:phi-equation} can be rewritten using the $\tr$- and  $\vartheta$-equations \eqref{eqn:r-equation} and \eqref{eqn:theta-equation}
\begin{equation}
	  \dd \varphi = \frac{\ta}{\tS}\left[ \left(\tr^2 -2\tb\tr +\ta^2  \right) E-\ta\tL \right]\frac{\dd\tr}{\sqrt{R}} - \frac{1}{\sin^2\!\vartheta} \left(\ta E \sin^2\!\vartheta -\tL\right) \frac{\dd\vartheta}{\sqrt{\Theta}} \, .
\end{equation}
Integrating this equation gives an  $\tr$-dependent integral $I_{\tr}$ and a $\vartheta$-dependent integral $I_\vartheta$ which can be solved separately
\begin{equation}
	  \varphi -  \varphi_0 =  \int_{\tr_{\rm in}}^{\tr} \! \frac{\ta}{\tS}\left[ \left(\tr^2 -2\tb\tr +\ta^2  \right) E-\ta\tL \right]\frac{\dd\tr}{\sqrt{R}} -  \int_{\vartheta_{\rm in}}^{\vartheta} \! \frac{1}{\sin^2\!\vartheta}\left(\ta E \sin^2\!\vartheta -\tL\right) \frac{\dd\vartheta}{\sqrt{\Theta}}  = I_{\tr} - I_\vartheta \, .
\end{equation}
First, let us consider the integral  $I_{\tr}$ and substitute  $\tr=\pm\frac{b_3}{4y-\frac{b_2}{3}}+\tr_R$. A partial fraction decomposition yields
\begin{equation}
	I_{\tr}= \int_{y_{\rm in}}^{y} \! \left( C_0 + \sum^2_{i=1}\frac{C_i}{y'-p_i} \right) \frac{\dd y'}{\sqrt{P^{\tr}_3(y')}} \, ,
\end{equation}
where the first order poles of $I_{\tr}$ are $p_1=\frac{b_2(\tr_+-\tr_R)\pm b_3}{12(\tr_+-\tr_R)}$ and $p_2=\frac{b_2(\tr_--\tr_R)\pm b_3}{12(\tr_--\tr_R)}$, which correspond to the horizons. The sign of $\pm b_3$ is determined by the chosen sign in the substitution $\tr=\pm\frac{b_3}{4y-\frac{b_2}{3}}+\tr_R$. $C_i$ are constants from the  partial fraction decomposition depending on the parameters of the metric and the test particle. A further substitution $y=\wp\left(\gamma - \gamma'_{\rm in}; g_2^{\tr}, g_3^{\tr}\right)=:\wp_{\tr}(v)$ with $v=\gamma-\gamma_{\rm in}'$ results in
\begin{equation}
	I_{\tr}= \int_{v_{\rm in}}^{v} \! \left( C_0 + \sum^2_{i=1}\frac{C_i}{\wp_{\tr}(v')-p_i} \right) \dd v' \, ,
\end{equation}
Analogously the integral $I_\vartheta$ can be brought into this specific form with the substitutions $\nu=\cos^2\vartheta$, then $\nu=\frac{1}{c_3}\left(4u-\frac{c_2}{3}\right)$ and lastly $u=\wp(\gamma-\gamma''_{\rm in};g_2^\vartheta,g_3^\vartheta)=:\wp_{\vartheta}(\tv)$ with $\tv=\gamma-\gamma_{\rm in}''$ 
\begin{equation}
	I_{\vartheta}= \int_{\tv_{\rm in}}^{\tv} \! \left( \ta E +\frac{c_3\tL}{4}\frac{1}{\wp_{\vartheta}(\tv')-q} \right) \dd \tv' \, ,
\end{equation}
where $q=\frac{c_3}{4}+\frac{c_2}{12}$ is a first order pole.

$I_{\tr}$ and $I_\vartheta$ are elliptic integrals of the third kind and can be solved with the help of the $\wp$-, $\sigma$- and $\zeta$-functions as shown in \cite{Kagramanova:2010bk, Grunau:2010gd, Enolski:2011id}. Finally, the solution of the $\varphi$-equation \eqref{eqn:phi-equation} is
\begin{equation}
	\begin{split}
		\varphi(\gamma) &= C_0(v-v_{\rm in}) + \sum^2_{i=1}\frac{C_i}{\wp'_{\tr}(v_i)}\left( 2\zeta_{\tr}(v_i)(v-v_{\rm in}) + \ln\frac{\sigma_{\tr}(v-v_i)}{\sigma_{\tr}(v_{\rm in}-v_i)} - \ln\frac{\sigma_{\tr}(v+v_i)}{\sigma_{\tr}(v_{\rm in}+v_i)}\right) \\
		&-\ta E(\tv-\tv_{\rm in}) - \frac{c_3\tL}{4\wp'_{\vartheta}(\tv_q)}\left( 2\zeta_{\vartheta}(\tv_q)(\tv-\tv_{\rm in}) + \ln\frac{\sigma_{\vartheta}(\tv-\tv_q)}{\sigma_{\vartheta}(\tv_{\rm in}-v_q)} - \ln\frac{\sigma_{\vartheta}(\tv+\tv_q)}{\sigma_{\vartheta}(\tv_{\rm in}+\tv_q)}\right) + \varphi_{\rm in}
	 \end{split}
\end{equation}
with $p_i=\wp_{\tr}(v_i)$, $q=\wp_{\vartheta}(\tv_q)$, $v=\gamma-\gamma_{\rm in}'$,  $\tv=\gamma-\gamma_{\rm in}''$  and
\begin{eqnarray}
	\wp_{\tr}(v) &= \wp (v, g_2^{\tr}, g_3^{\tr})\, , \qquad \wp_\vartheta (\tv)&= \wp (\tv, g_2^{\vartheta}, g_3^{\vartheta}) \, ,\nonumber\\
	\zeta_{\tr}(v) &= \zeta (v, g_2^{\tr}, g_3^{\tr})\, , \qquad \zeta_\vartheta (\tv)&= \zeta (\tv, g_2^{\vartheta}, g_3^{\vartheta}) \, ,\\
	\sigma_{\tr}(v) &= \sigma (v, g_2^{\tr}, g_3^{\tr})\, , \qquad \sigma_\vartheta (\tv)&= \sigma (\tv, g_2^{\vartheta}, g_3^{\vartheta}) \, .\nonumber
\end{eqnarray}

\subsection{The $\tlt$-equation}

The $\tlt$-equation \eqref{eqn:t-equation} can be rewritten using the $\tr$- and  $\vartheta$-equations \eqref{eqn:r-equation} and \eqref{eqn:theta-equation}
\begin{equation}
	\dd \tlt = \frac{\tr^2-2\tb\tr+\ta^2}{\tS}\left[\left(\tr^2-2\tb\tr+\ta^2\right)E-\ta\tL\right] \frac{\dd\tr}{\sqrt{R}} - \ta\left(\ta E \sin^2\!\vartheta -\tL\right) \frac{\dd\vartheta}{\sqrt{\Theta}} \, .
\label{eqn:tsplit}
\end{equation}
Integrating this equation gives an  $\tr$-dependent integral and a $\vartheta$-dependent integral as in the case of the $\varphi$-equation \eqref{eqn:phi-equation}
\begin{equation}
	\tlt - \tlt_{\rm in} = \int_{v_{\rm in}}^{v} \! \left( C'_0 + \sum^2_{i=1}\frac{C'_i}{\wp_{\tr}(v')-p_i} 
	\right) \dd v' 
	- \int_{\tv_{\rm in}}^{\tv} \! \left( \ta^2E-\ta\tL+\frac{c_2}{3} - \frac{4\ta^2E}{c_3}\wp_{\vartheta}(\tv) \right) \dd \tv' \, ,
\end{equation}
with the  first order poles $p_1=\frac{b_2(\tr_+-\tr_R)\pm b_3}{12(\tr_+-\tr_R)}$ and $p_2=\frac{b_2(\tr_--\tr_R)\pm b_3}{12(\tr_--\tr_R)}$. Here we used the same substitutions as in section \ref{sec:phisol} and applied a partial fraction decomposition which gave rise to the constants $C'_i$. Again we find elliptic integrals of the third kind that can be solved with the help of the $\wp$-, $\sigma$- and $\zeta$-functions as shown in \cite{Kagramanova:2010bk, Grunau:2010gd, Enolski:2011id}.Finally, the solution of the $\tlt$-equation \eqref{eqn:t-equation} is
\begin{equation}
	\begin{split}
		\tlt(\gamma) &= C'_0(v-v_{\rm in}) + \sum^2_{i=1}\frac{C'_i}{\wp'_{\tr}(v_i)}\left( 2\zeta_{\tr}(v_i)(v-v_{\rm in}) + \ln\frac{\sigma_{\tr}(v-v_i)}{\sigma_{\tr}(v_{\rm in}-v_i)} - \ln\frac{\sigma_{\tr}(v+v_i)}{\sigma_{\tr}(v_{\rm in}+v_i)}\right) \\
		& -\left( \ta^2E-\ta\tL+\frac{c_2}{3}\right) (\tv-\tv_{\rm in}) -  \frac{4\ta^2E}{c_3}(\zeta_{\vartheta}(\tv)-\zeta_{\vartheta}(\tv_{\rm in}))+ \tlt_{\rm in} \, ,
	\end{split}
\end{equation}
with $p_i=\wp_{\tr}(v_i)$, $v=\gamma-\gamma_{\rm in}'$,  $\tv=\gamma-\gamma_{\rm in}''$  and
\begin{eqnarray}
	\wp_{\tr}(v) &= \wp (v, g_2^{\tr}, g_3^{\tr})\, , \qquad \wp_\vartheta (\tv)&= \wp (\tv, g_2^{\vartheta}, g_3^{\vartheta}) \, ,\nonumber\\
	\zeta_{\tr}(v) &= \zeta (v, g_2^{\tr}, g_3^{\tr})\, , \qquad \zeta_\vartheta (\tv)&= \zeta (\tv, g_2^{\vartheta}, g_3^{\vartheta}) \, ,\\
	\sigma_{\tr}(v) &= \sigma (v, g_2^{\tr}, g_3^{\tr})\, , \qquad \sigma_\vartheta (\tv)&= \sigma (\tv, g_2^{\vartheta}, g_3^{\vartheta}) \, .\nonumber
\end{eqnarray}

\section{The orbits}

\begin{figure}[p]
	\centering
	\subfigure[$\delta=1$, $\ta=0.4$, $\tL=-0.5$, $\tK=5$, ${\tb=-0.08}$, ${E=0.97}$: Bound orbit]{
		\includegraphics[width=0.31\textwidth]{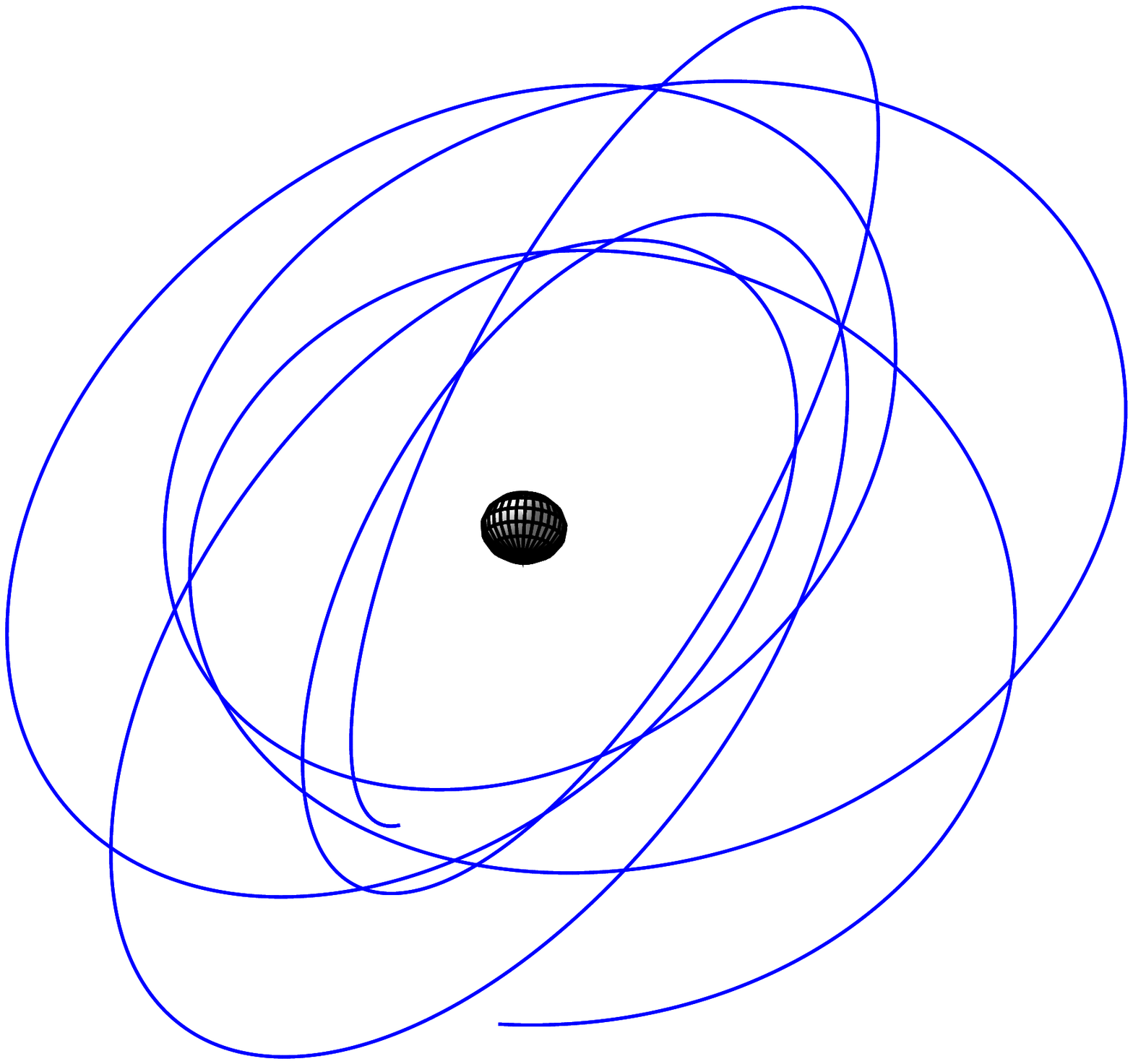}
	}
	\subfigure[$\delta=1$, $\ta=0.4$, $\tL=0.8$, $\tK=2$, ${\tb=-0.08}$, ${E=1.002}$: Escape orbit]{
		\includegraphics[width=0.31\textwidth]{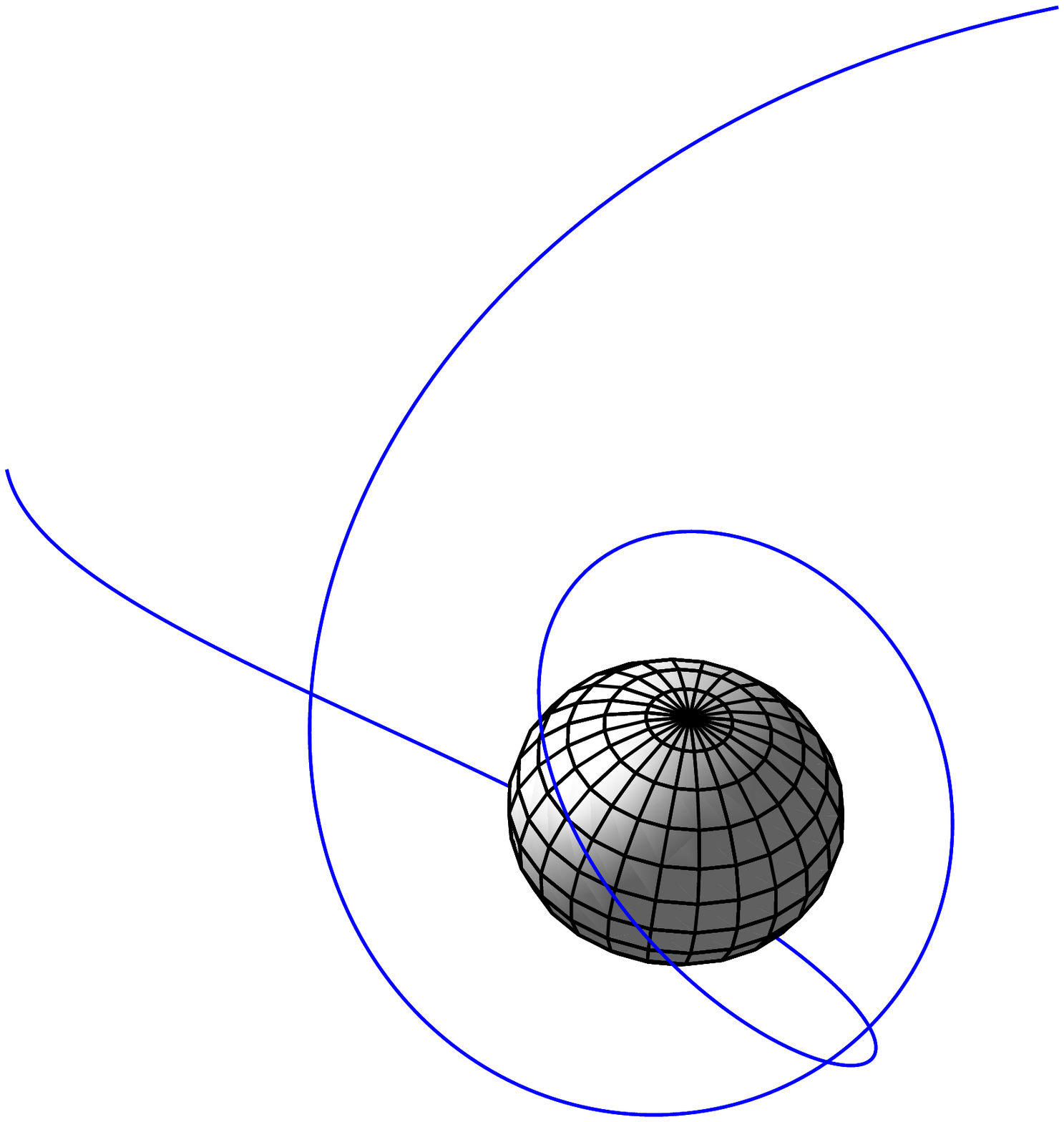}
	}
	\subfigure[$\delta=1$, $\ta=0.4$, $\tL=1$, $\tK=0.36$, ${\tb=-0.08}$, ${E=4}$: Terminating orbit]{
		\includegraphics[width=0.31\textwidth]{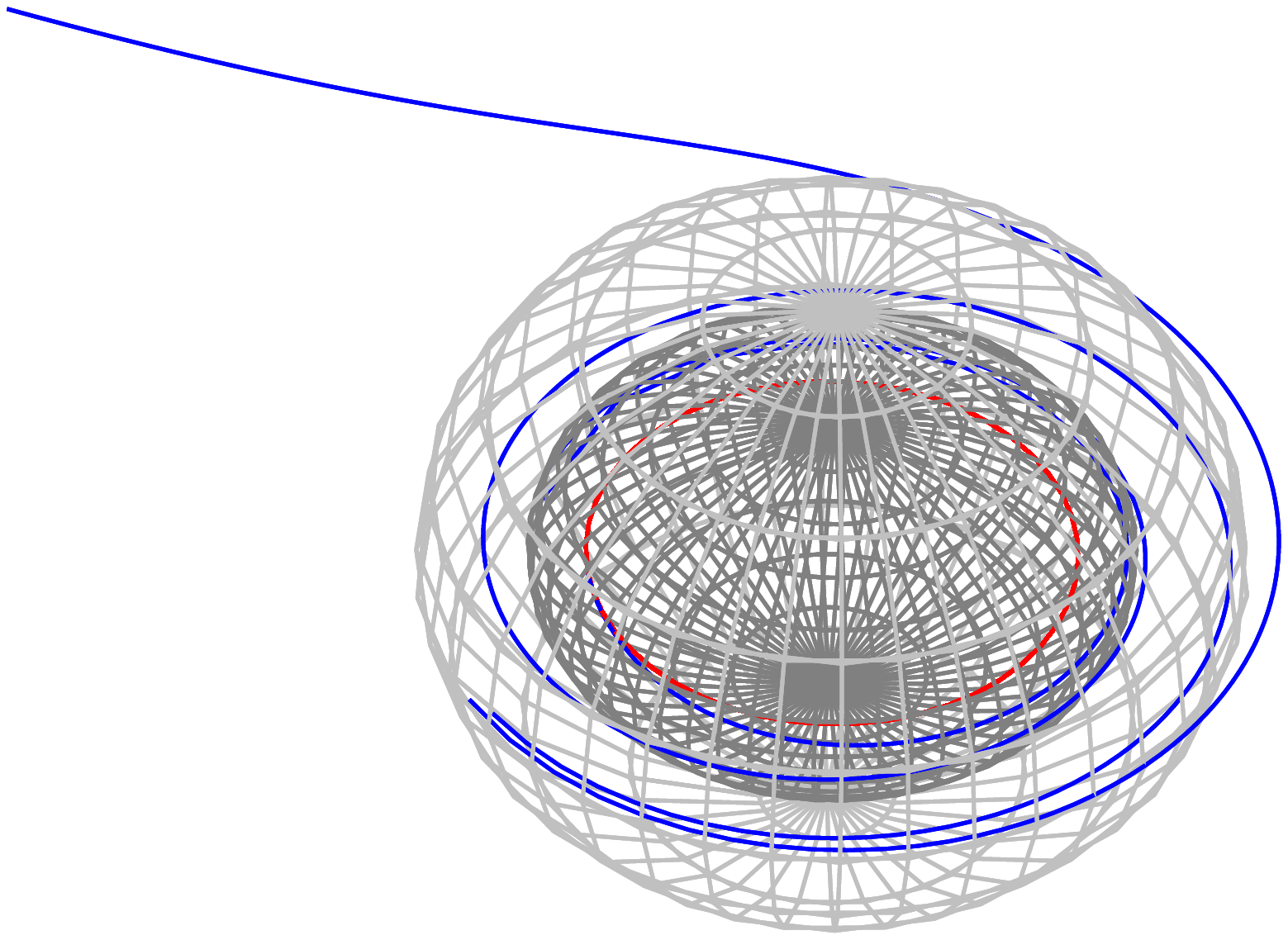}
	}
	\caption{Different orbits of test particles in the EMDA black hole spacetime. The blue lines represent the orbits and the ellipsoids indicate the positions of the outer horizon $\tr_+$ and the inner horizon $\tr_-$. The red circle in figure (c) is the singularity at ${\tr=0}$ and ${\vartheta=\frac{\pi}{2}}$.}
 \label{pic:orbits-1}
\end{figure}

\begin{figure}[p]
	\centering
	\subfigure[$\delta=1$, $\ta=0.2$, $\tL=1$, $\tK=0.008$, ${\tb=-0.08}$, ${E=4.7}$: Bound orbit with $\tr<0$]{
		\includegraphics[width=0.31\textwidth]{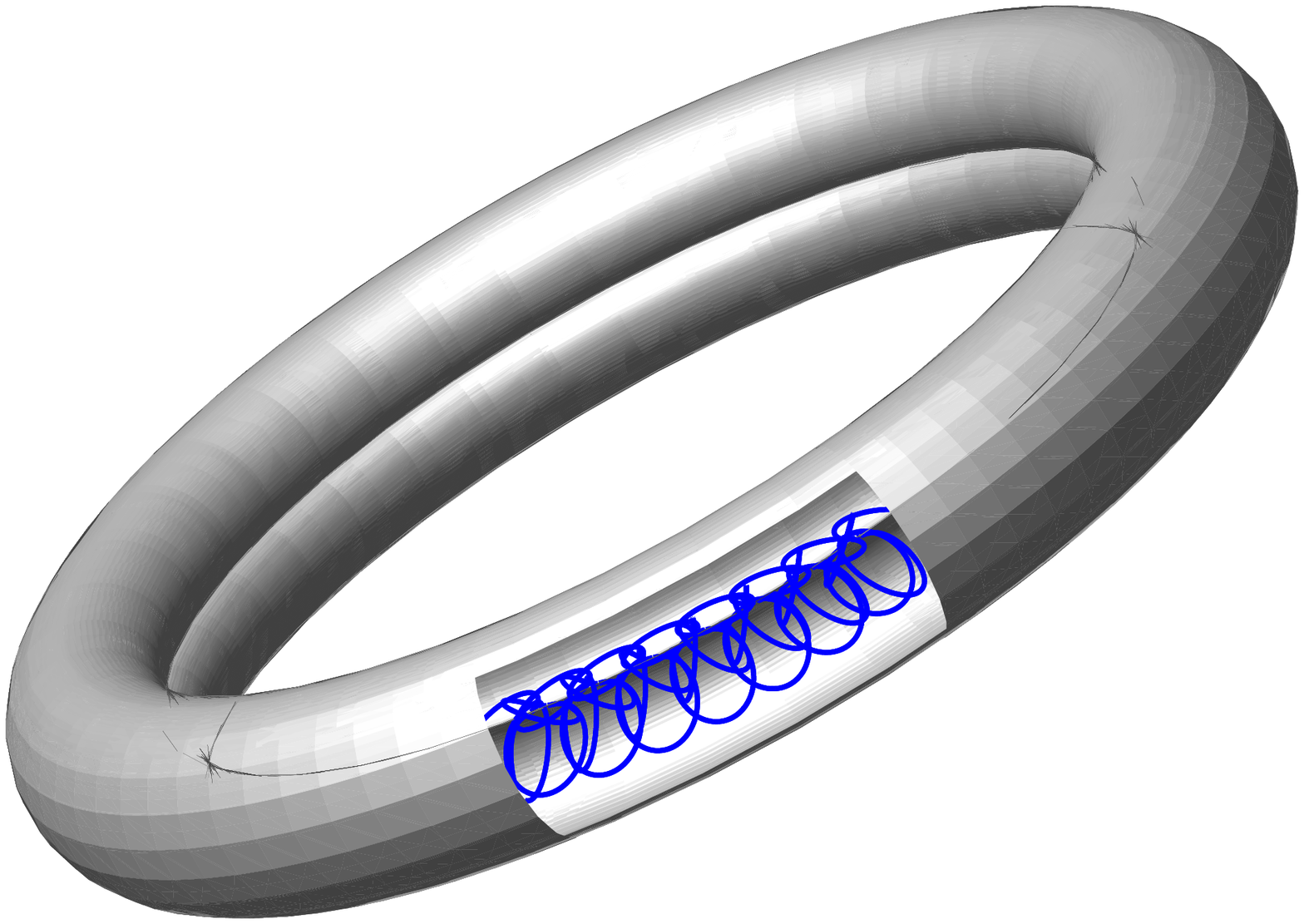}
	}
	\subfigure[$\delta=1$, $\ta=0.078$, $\tL=0.15$, $\tK=5$, ${\tb=-0.08}$, ${E=1.1}$: Bound orbit orbit with $\tr<0$]{
		\includegraphics[width=0.31\textwidth]{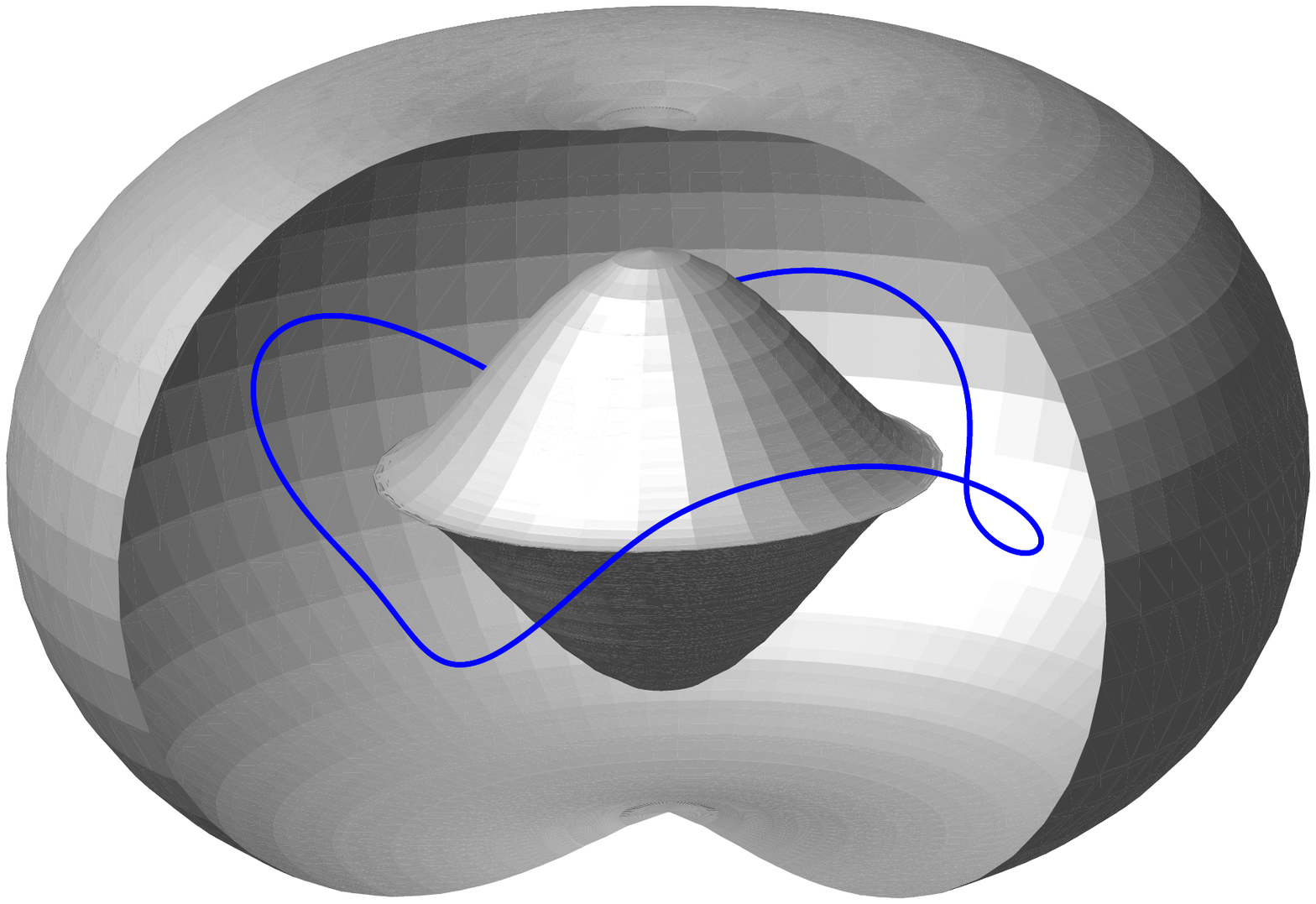}
	}
	\subfigure[$\delta=1$, $\ta=0.35$, $\tL=1$, $\tK=2$, ${\tb=-0.08}$, ${E=8.5}$: Escape orbit with $\tr<0$]{
		\includegraphics[width=0.31\textwidth]{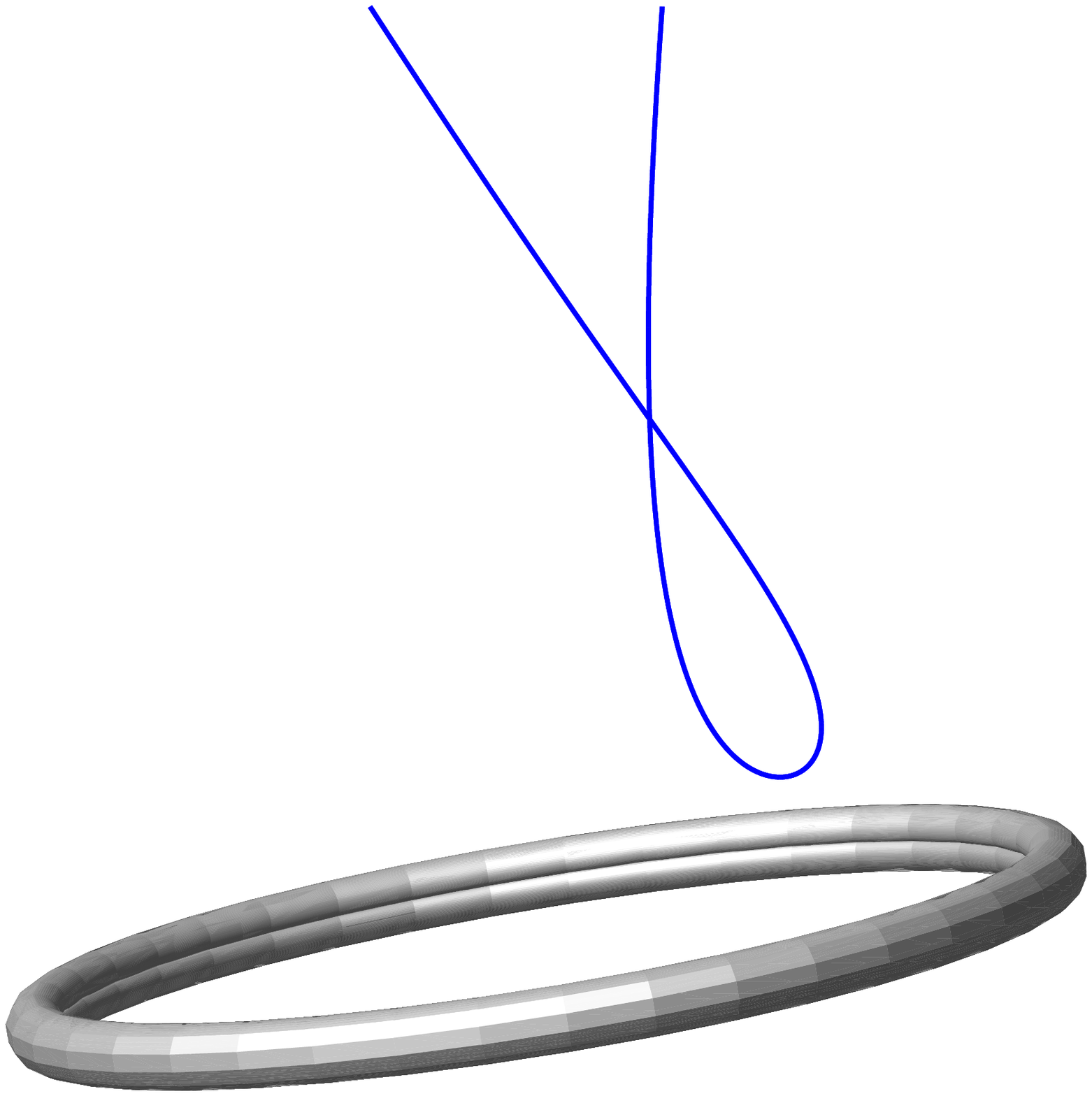}
	}
	\caption{Different orbits of test particles in the EMDA black hole spacetime. For the orbits depicted here, the coordinate $\tr$ is negative and the singularity is not covered by a horizon. The grey structure is the singularity (compare section \ref{sec:EMDA}, figure \ref{pic:singularity-shape}). In figure (a) and (b), we removed a part of the singularity from the plot, so that the orbit inside becomes visible.}
 \label{pic:orbits-3}
\end{figure}

In this section we use the analytical solutions of the equations of motion to plot some examples of test particle orbits in the EMDA black hole spacetime, see figure \ref{pic:orbits-1}--figure \ref{pic:orbits-3}. A bound orbit and an escape orbit are depicted in figure \ref{pic:orbits-1}(a) and figure \ref{pic:orbits-1}(b) respectively. A terminating orbit, which lies in the equatorial plane, is shown in figure \ref{pic:orbits-1}(c). There one can also observe that the orbit (blue line) is discontinuous when it crosses a horizon. This is due to divergencies in the $\varphi$-equation, which occur at the horizons.

\begin{figure}[p]
	\centering
	\subfigure[$\delta=1$, $\ta=0.4$, $\tL=1$, $\tK=0.5$, ${\tb=-0.08}$, ${E=5.5}$: Two-world escape orbit]{
		\includegraphics[width=0.31\textwidth]{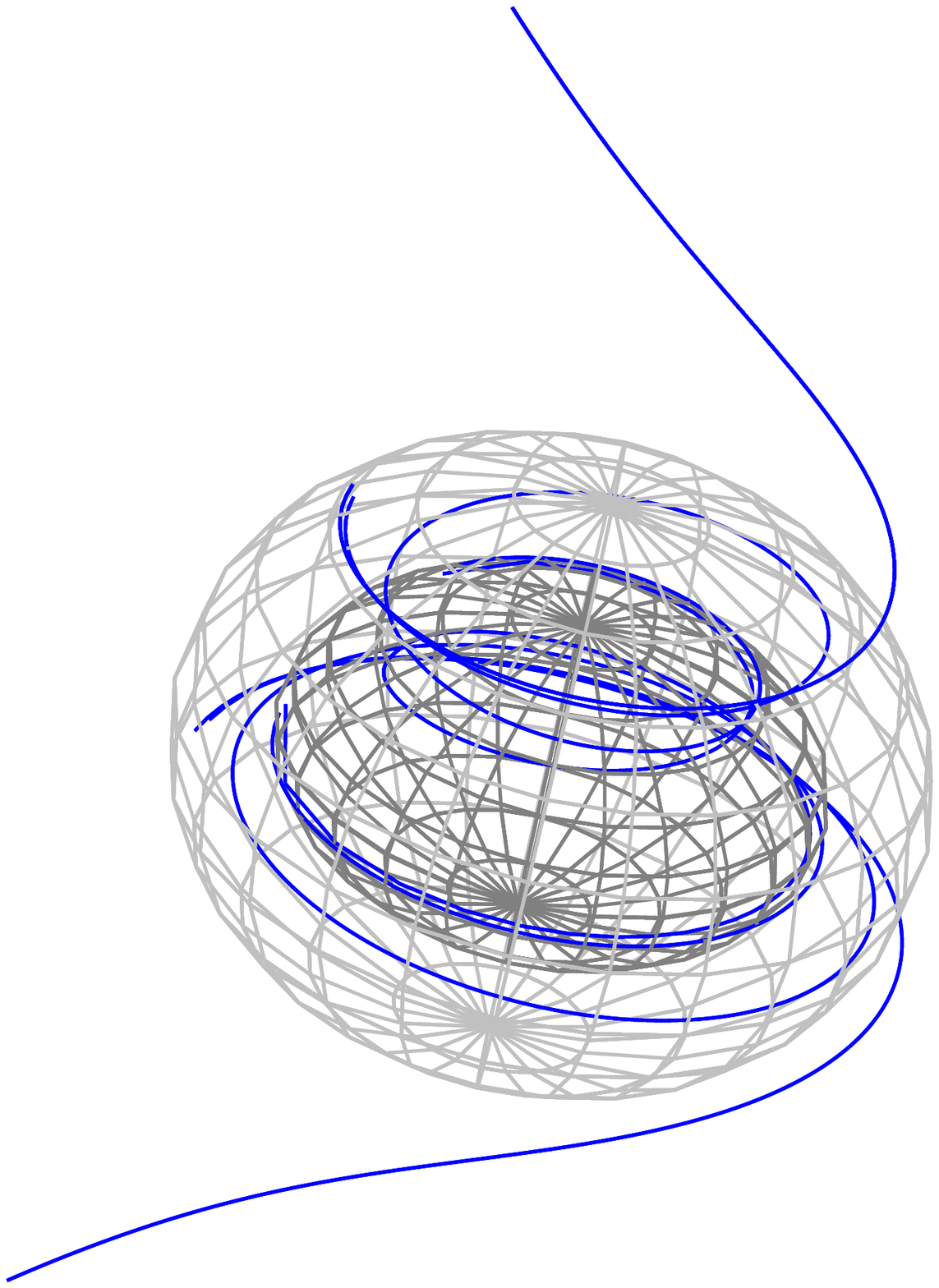}
	}
	\subfigure[$\delta=1$, $\ta=0.4$, $\tL=-1$, $\tK=5$, ${\tb=-0.08}$, ${E=0.8}$: Many-world bound orbit]{
		\includegraphics[width=0.31\textwidth]{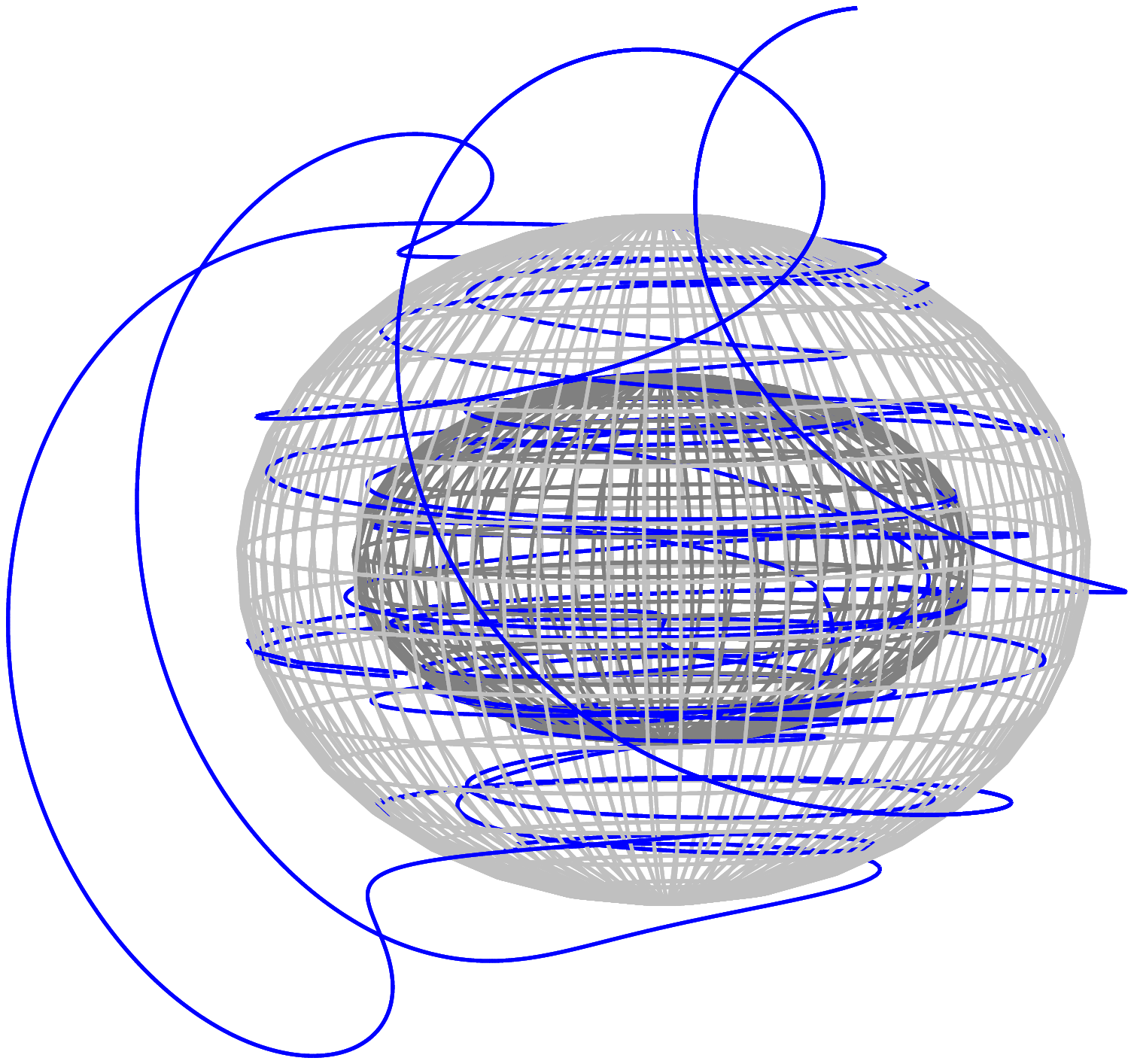}
	}
	\subfigure[$\delta=1$, $\ta=0.4$, $\tL=0.5$, $\tK=1$, ${\tb=-0.08}$, ${E=5}$: Transit orbit]{
		\includegraphics[width=0.31\textwidth]{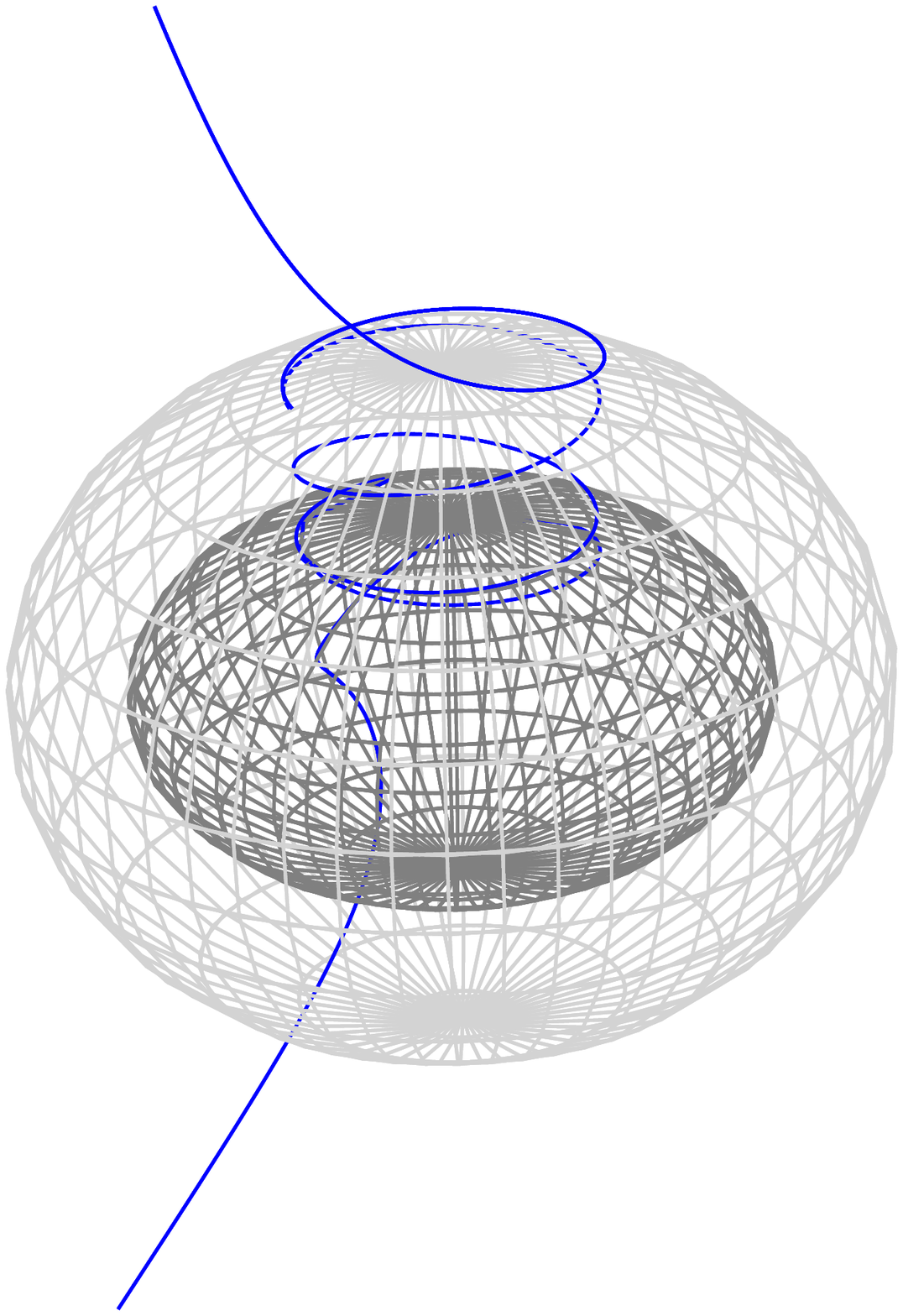}
	}
	\caption{Different orbits of test particles in the EMDA black hole spacetime. The blue lines represent the orbits and the ellipsoids indicate the positions of the outer horizon $\tr_+$ and the inner horizon $\tr_-$. }
 \label{pic:orbits-2}
\end{figure}

Figure \ref{pic:orbits-2} shows three orbits that cross both horizons, but do not end in the singularity. On the two-world escape orbit (figure \ref{pic:orbits-2}(a)) and the many-world bound orbit (\ref{pic:orbits-2}(b)) the test particles emerge into another universe each time the horizons are crossed twice. In figure \ref{pic:orbits-2}(b) the effect of the ergosphere is visible, since here the angular momentum of the black hole and the angular momentum of the particle have opposite signs. If the test particle enters the ergosphere, it is forced to rotate along with the black hole and therefore changes its direction. The particle on the bound orbit in figure \ref{pic:orbits-1}(a) has also an angular momentum with a different sign than the rotation of the black hole, but this motion takes place far away from the ergosphere and maintains its direction.

The test particle on the transit orbit depicted in \ref{pic:orbits-2}(c) crosses $\tr=0$ (but not the equatorial plane) and travels into a universe where gravity acts repulsive. Therefore in the orbit plot it looks like the particle is reflected at that point.

In figure \ref{pic:orbits-3} two bound orbits (figure \ref{pic:orbits-3}(a) and \ref{pic:orbits-3}(b)) and an escape orbit  (figure \ref{pic:orbits-3}(c)) with negative $\tr$ around a naked singularity can be seen. The singularity at negative $\tr$ acquires different shapes depending on the parameters of the metric; see also section \ref{sec:EMDA}. Bound orbits at $\tr<0$ are always hidden inside the ``walls'' of the singularity. This region enclosed by the surfaces of the singularity can never be reached from the outside, nor can a particle leave the inside, since the singularity is repulsive here. This is also observed in the plots of the effective potential in figure \ref{pic:potentials}(e) and  \ref{pic:potentials}(f), which are similar to the corresponding potential of the bound orbit shown in \ref{pic:orbits-3}(b). Although the region inside the singularity can never be reached from the outside, we display the orbits for the sake of completeness.


\section{Conclusion}

In this article we studied the EMDA spacetime with the help of geodesics and presented the analytical solutions of the equations of motion. The geodesic equations are of elliptic type and can be solved in terms of the Weierstra{\ss} $\wp$-, $\sigma$- and $\zeta$-functions. The dilaton parameter $\tb$ influences the $\tr$-equation, $\varphi$-equation and $\tlt$-equation, but the $\vartheta$-equation is not affected and is exactly the same as in the Kerr spacetime, which is recovered by setting $\tb=0$. The dilaton charge changes the shape of the singularity. At $\tr=0$ and $\vartheta=\frac{\pi}{2}$ we still have the ring singularity, but for negative $\tr$ the singularity forms a toruslike structure or two closed surfaces depending on the dilaton charge and the rotation parameter. Similar shapes of the singularity of a black hole in Einstein-Maxwell-dilaton-axion theory have been observed in \cite{Matos:2009rp}. Since gravity becomes repulsive for $\tr<0$, geodesics can only hit the singularity at $\tr=0$ and $\vartheta=\frac{\pi}{2}$ (as in the Kerr spacetime, these geodesics lie in the equatorial plane).

To determine the possible orbits in the EMDA spacetime, parametric $\tL$-$E^2$-diagrams and effective potential techniques were used. We found terminating orbits that end in the singularity and lie in the equatorial plane, escape orbits and bound orbits. In this spacetime escape orbits are possible both for positive and negative $\tr$. Bound orbits occur for $\tr>\tr_+$ and for $0< \tr <\tr_-$ hidden behind the inner horizon, but also for $\tr<0$. These bound orbits at negative $\tr$ do not occur in the Kerr case $\beta\neq0$ and are hidden inside the ``walls'' of the singularity. Other possible orbits are two-world escape orbits and many-world bound orbits which emerge into other universes every time the horizons are crossed twice. Additionally it is possible to cross $\tr=0$ and travel into an antigravity universe, resulting in transit orbits with the range $-\infty < \tr < \infty$ and crossover two-world escape orbits which cross both the horizons and $\tr=0$.

The exact orbits and their properties can be calculated using the analytical solutions of the geodesic equations. Additionally, observables like the periastron shift of a bound orbit or the light deflection of an escape orbit can derived with formulas analogous to \cite{Hackmann:2010zz}. The geodesics of photons can be used to calculate the shadow of the black hole. In the EMDA spacetime the shadow was studied numerically in \cite{Wei:2013kza}. Hopefully one day all this could be compared to observations.\\

For future work it might be interesting to extend the equations of motion and their solutions to electrically and magnetically charged particles. Then the influence of the dilaton on the charged particles could be studied and compared to the particle motion in e.g. the Kerr-Newman spacetime. The analytical solutions to the geodesic equations of charged particles in the Kerr-Newman spacetime were found in \cite{Hackmann:2013pva}. Moreover, the NUT parameter could be included by setting $b\neq 0$ in the spacetime of \cite{Garcia:1995qz}. Another option is to consider particle motion around a more general black hole spacetime containing more charges and a cosmological constant. An interesting candidate for this is the dyonic AdS black hole in maximal gauged supergravity found by Chow and Comp\`{e}re \cite{Chow:2013gba}. The EMDA black hole studied in the present paper represents a special case of the spacetime in \cite{Chow:2013gba}.

\section{Acknowledgements}
We would like to thank Jutta Kunz for interesting and fruitful discussions. 
S.G. gratefully acknowledges support by the DFG within the Research Training Group 1620 ``Models of Gravity''.


\bibliographystyle{unsrt}

\begin{thebibliography}{99}

\bibitem{Matarrese:2011}
  S.~Matarrese, M.~Colpi, V.~Gorini and ~U. Moschella (Eds.)
  Dark Matter and Dark Energy,
  Astrophysics and Space Science Library, Vol. 370,
  Springer 2011

\bibitem{Garcia:1995qz} 
  A.~Garcia, D.~Galtsov and O.~Kechkin,
  Phys.\ Rev.\ Lett.\  {\bf 74}, 1276 (1995).

\bibitem{Hagihara:1931}
 Y.~Hagihara,
 Jpn.\ J.\ Astron.\ Geophys.\ {\bf 8}, 67 (1931)

\bibitem{Kagramanova:2010bk} 
  V.~Kagramanova, J.~Kunz, E.~Hackmann and C.~L\"ammerzahl,
  Phys.\ Rev.\ D {\bf 81}, 124044 (2010)
  [arXiv:1002.4342 [gr-qc]].

\bibitem{Grunau:2010gd} 
  S.~Grunau and V.~Kagramanova,
  Phys.\ Rev.\ D {\bf 83}, 044009 (2011)
  [arXiv:1011.5399 [gr-qc]].

\bibitem{Kagramanova:2012hw} 
  V.~Kagramanova and S.~Reimers,
  Phys.\ Rev.\ D {\bf 86}, 084029 (2012)
  [arXiv:1208.3686 [gr-qc]].

\bibitem{Hackmann:2013pva} 
  E.~Hackmann and H.~Xu,
  Phys.\ Rev.\ D {\bf 87}, no. 12, 124030 (2013)
  [arXiv:1304.2142 [gr-qc]].

\bibitem{Hackmann:2008zza} 
  E.~Hackmann and C.~L\"ammerzahl,
  Phys.\ Rev.\ Lett.\  {\bf 100}, 171101 (2008).

\bibitem{Hackmann:2008zz} 
  E.~Hackmann and C.~L\"ammerzahl,
  Phys.\ Rev.\ D {\bf 78}, 024035 (2008).

\bibitem{Hackmann:2008tu} 
  E.~Hackmann, V.~Kagramanova, J.~Kunz and C.~L\"ammerzahl,
  Phys.\ Rev.\ D {\bf 78}, 124018 (2008)
  [Erratum-ibid.\  {\bf 79}, 029901 (2009)]
  [arXiv:0812.2428 [gr-qc]].

\bibitem{Hackmann:2010zz} 
  E.~Hackmann, C.~L\"ammerzahl, V.~Kagramanova and J.~Kunz,
  Phys.\ Rev.\ D {\bf 81}, 044020 (2010)
  [arXiv:1009.6117 [gr-qc]].

\bibitem{Enolski:2010if} 
  V.~Z.~Enolski, E.~Hackmann, V.~Kagramanova, J.~Kunz and C.~L\"ammerzahl,
  J.\ Geom.\ Phys.\  {\bf 61}, 899 (2011)
  [arXiv:1011.6459 [gr-qc]].

\bibitem{Enolski:2011id} 
  V.~Enolski, B.~Hartmann, V.~Kagramanova, J.~Kunz, C.~L\"ammerzahl and P.~Sirimachan,
  Journal of mathematical physics {\bf{53}}, 012504 (2012)  
  [arXiv:1106.2408 [gr-qc]].

\bibitem{Grunau:2012ai} 
  S.~Grunau, V.~Kagramanova, J.~Kunz and C.~L\"ammerzahl,
  Phys.\ Rev.\ D {\bf 86}, 104002 (2012)
  [arXiv:1208.2548 [gr-qc]].

\bibitem{Grunau:2012ri} 
  S.~Grunau, V.~Kagramanova and J.~Kunz,
  Phys.\ Rev.\ D {\bf 87}, 044054 (2013)
  arXiv:1212.0416 [gr-qc].

\bibitem{Aliev:1988wv} 
  A.~N.~Aliev and D.~V.~Galtsov,
  Sov.\ Astron.\ Lett.\  {\bf 14}, 48 (1988).

\bibitem{Galtsov:1989ct} 
  D.~V.~Galtsov and E.~Masar,
  Class.\ Quant.\ Grav.\  {\bf 6}, 1313 (1989).

\bibitem{Chakraborty:1991mb} 
  S.~Chakraborty and L.~Biswas,
  Class.\ Quant.\ Grav.\  {\bf 13}, 2153 (1996).

\bibitem{Ozdemir:2003km} 
  N.~Ozdemir,
  Class.\ Quant.\ Grav.\  {\bf 20}, 4409 (2003).

\bibitem{Ozdemir:2004ne} 
  F.~Ozdemir, N.~Ozdemir and B.~T.~Kaynak,
  Int.\ J.\ Mod.\ Phys.\ A {\bf 19}, 1549 (2004).

\bibitem{Grunau:2013oca} 
  S.~Grunau and B.~Khamesra,
  Phys.\ Rev.\ D {\bf 87}, no. 12, 124019 (2013)
  [arXiv:1303.6863 [gr-qc]].

\bibitem{Hartmann:2010rr} 
  B.~Hartmann and P.~Sirimachan,
  JHEP {\bf 1008}, 110 (2010)
  [arXiv:1007.0863 [gr-qc]].

\bibitem{Hartmann:2012pj} 
  B.~Hartmann and V.~Kagramanova,
  Phys.\ Rev.\ D {\bf 86}, 045028 (2012)
  [arXiv:1204.0396 [hep-th]].

\bibitem{Hartmann:2010vp} 
  B.~Hartmann, C.~L\"ammerzahl and P.~Sirimachan,
  Phys.\ Rev.\ D {\bf 83}, 045027 (2011)
  [arXiv:1012.3285 [hep-th]].

\bibitem{Hackmann:2009rp} 
  E.~Hackmann, B.~Hartmann, C.~L\"ammerzahl and P.~Sirimachan,
  Phys.\ Rev.\ D {\bf 81}, 064016 (2010)
  [arXiv:0912.2327 [gr-qc]].

\bibitem{Hackmann:2010ir} 
  E.~Hackmann, B.~Hartmann, C.~L\"ammerzahl and P.~Sirimachan,
  Phys.\ Rev.\ D {\bf 82}, 044024 (2010)
  [arXiv:1006.1761 [gr-qc]].

\bibitem{Carter:1968rr} 
  B.~Carter,
  Phys.\ Rev.\  {\bf 174}, 1559 (1968).

\bibitem{Mino:2003yg} 
  Y.~Mino,
  Phys.\ Rev.\ D {\bf 67}, 084027 (2003)
  [gr-qc/0302075].

\bibitem{Markushevich:1967}
  A.~I.~Markushevich, {\em Theory of Functions of a Complex Variable} (Prentice-Hall, Englewood Cliffs, NJ, 1967), Vol. III.

\bibitem{ONeill:1995}
 B.~O'Neill, {\em The Geometry of Kerr Black Holes} (AK Peters, Wellesley, Massachusetts, 1995).

\bibitem{Matos:2009rp} 
  T.~Matos, G.~Miranda, R.~Sanchez-Sanchez and P.~Wiederhold,
  Phys.\ Rev.\ D {\bf 79}, 124016 (2009)
  [arXiv:0905.4097 [gr-qc]].

\bibitem{Wei:2013kza} 
  S.~W.~Wei and Y.~X.~Liu,
  JCAP {\bf 1311}, 063 (2013)
  [arXiv:1311.4251 [gr-qc]].

\bibitem{Chow:2013gba} 
  D.~D.~K.~Chow and G.~Comp\`{e}re,
  Phys.\ Rev.\ D {\bf 89}, no. 6, 065003 (2014)
  [arXiv:1311.1204 [hep-th]].

\end{thebibliography}

\end{document}